\documentclass[9pt]{article}  \usepackage{times}
\usepackage{graphicx, textgreek}

\usepackage{color}

\usepackage[round,numbers,sort&compress]{natbib} 

\usepackage{amsmath}
\usepackage{amssymb}
\usepackage{stackrel}
\usepackage{chemarrow}
\usepackage{graphicx}
\usepackage{textgreek}

\usepackage{booktabs}
\usepackage{dcolumn}
\usepackage{rotating}
\usepackage{multirow}

\renewcommand\appendix{\par
	\setcounter{section}{0}
	\setcounter{subsection}{0}
	\setcounter{figure}{0}
	\setcounter{table}{0}
	\renewcommand\thesection{Appendix \Alph{section}}
	\renewcommand\thefigure{\Alph{section}\arabic{figure}}
	\renewcommand\thetable{\Alph{section}\arabic{table}}
}

\begin{document}

\twocolumn[{\LARGE \textbf{The important consequences of the reversible heat production in nerves and the adiabaticity of the action potential\\*[0.2cm]}}
{\large Thomas Heimburg$^\ast$\\*[0.1cm]
	{\small Niels Bohr Institute, University of Copenhagen, Blegdamsvej 17, 2100 Copenhagen \O, Denmark}\\*[-0.1cm]
	
	{\normalsize \textbf{ABSTRACT}\hspace{0.5cm} It has long been known that there is no measurable heat production associated with the nerve pulse. Rather, one finds that heat production is biphasic, and a heat release during the first phase of the action potential is followed by the reabsorption of a similar amount of heat during the second phase. We review the long history the measurement of heat production in nerves and provide a new analysis of these findings focusing on the thermodynamics of adiabatic and isentropic processes. We begin by considering adiabatic oscillations in gases, waves in layers, oscillations of springs and the reversible (or irreversible) charging and discharging of capacitors. We then apply these ideas to the heat signature of nerve pulses. Finally, we compare the temperature changes expected from the Hodgkin-Huxley model and the soliton theory for nerves.
		We demonstrate that heat production in nerves cannot be explained as an irreversible charging and discharging of a membrane capacitor as it is proposed in the Hodgkin-Huxley model. Instead, we conclude that it is consistent with an adiabatic pulse. However, if the nerve pulse is adiabatic, completely different physics is required to explain its features. Membrane processes must then be reversible and resemble the oscillation of springs more than resembling ``a burning fuse of gunpowder'' (quote A. L. Hodgkin). Theories acknowledging the adiabatic nature of the nerve pulse have recently been discussed by various authors. It forms the central core of the soliton model, which considers the nerve pulse as a localized sound pulse.
		\\*[0.3cm] }}
\noindent\footnotesize{\textbf{Keywords:} action potential; heat production; adiabaticity; sound; nerves\\*[0.1cm]}
\noindent\footnotesize {$^{\ast}$corresponding author, theimbu@nbi.ku.dk. }\\
\vspace{0.3cm}
]

\normalsize

\section{Introduction}
\label{introduction}

The nerve pulse is generally regarded as a purely electrical phenomenon \cite{Johnston1995}. However, one also finds measurable changes in nerve thickness, length and temperature. The latter changes are not well-known and, if they are recognized at all, are considered low-amplitude side-effects of the voltage pulse \cite{ElHady2015}. In the present paper we focus on the temperature changes and the remarkable fact that the magnitude of this signal displays a larger energy than that related to the voltage changes - i.e., it is not a small signal.

In 1845, Emil du Bois-Reymond was the first to measure electrical currents and voltage changes in stimulated muscles and nerves \cite{duBoisReymond1848}, which were found to be of similar magnitude. Hermann von Helmholtz was familiar with du Bois-Reymond's experiments. In 1852, he was the first to determine the velocity of the nervous impulse \cite{Helmholtz1852}, which he found to be close to the velocity of sound. Besides having a degree in physiology, Helmholtz was also an exquisite physicist and later became a professor of physics in Berlin. In 1847, Helmholtz proposed the first law of thermodynamics \cite{Helmholtz1847}. It states that the change of energy in a system is related to the absorption or release of heat and the work performed. Du Bois-Reymond's nerve and muscle experiments combined with his own considerations made Helmholtz interested in the heat production of muscles and nerves. Considering the similarities of their currents, he expected that chemical reactions and heat generation in muscles and nerves would also be similar. When performing experiments on frog muscles, he could measure temperature changes after repetitive muscle contraction that were of the order of 0.035$^\circ$C \cite{Helmholtz1848}. With the same experimental setup he found that the temperature changes in the nerves leading to this muscle were unmeasurably small (i.e., smaller than 0.002$^\circ$C, which was the sensitivity limit of his experiment). This finding indicated that the mechanisms of heat production in muscles and nerves must be very different. In particular, the lack of measurable heat production indicates that no metabolism took place during the nerve pulse and that the process under consideration might be of reversible physical nature. Helmholtz found this negative result interesting enough to dedicate a section of his paper to heat production in nerves \cite{Helmholtz1848}.

Stimulated by von Helmholtz' intriguing finding, similar experiments were performed in subsequent research while increasing the sensitivity of the experimental setup. In 1868, Heidenreich \cite{Heidenhain1868} improved the sensitivity of the heat recordings by about a factor of 10 using a thermopile. Like Helm\-holtz, he did not find any heat liberation during the nervous impulse. In 1890, Rolleston \cite[]{Rolleston1890} came to a similar conclusion using a platinum resistance thermometer. In 1891, Stewart \cite[]{Stewart1891} repeated the experiments on mammalian nerve using a similar setup. He also found no heat production. In 1897, Cremer \cite[]{Cremer1897} also failed to find any heat production in olfactory nerves of pike, carp, and barbel. In 1912, Archibald Vivian Hill \cite[]{Hill1912} performed experiments on sciatic nerves of frogs (Rana temporaria) using a very sensitive thermopile consisting of 30 iron-constatan thermocouples in series. He summarized:

\begin{quote}
	\emph{``By a thermo-electric method it is shown that tetanus, up to 25 secs., of a live nerve, does not cause a change of temperature (other than at the seat of excitation) of more than about $+6\cdot 10^{-6\;\circ}$C. There is no evidence of any change at all, but the method does not allow conclusions beyond this limit. For every single propagated disturbance the change of temperature therefore cannot exceed about $10^{-8\;\circ}$C., a hundred million$^{th}$ of a degree. This corresponds to an oxidative process, in which only one molecule of oxygen is used in a space of visible size, viz. a 3.7 \textmu cube. This suggests very strongly, though of course it does not finally prove, that the propagated nervous impulse is not a wave of irreversible chemical breakdown, but a reversible change of a purely physical nature.''}
\end{quote}

Later in his life \cite{Hill1959} he commented about this early phase of heat measurements:

\begin{quote}
	\emph{``Why did people go on trying to measure the heat production of nerve, in spite of repeated failure? Chiefly, I suppose, in order to settle the question of whether the nerve impulse is the sort of physical wave in which the whole of the energy for transmission is impressed on the system at the start. ''}
\end{quote}

W. M. Bayliss \cite{Bayliss1915} summarized the situation in his textbook `Principles of general physiology':

\begin{quote}
	\emph{``The result makes it impossible to suppose that any chemical process resulting in an irreversible loss of energy can be involved in the transmission of a nerve impulse, and indicates that a reversible physicochemical one of some kind is to be looked for.''}
\end{quote}

Thus, the conclusion that had to be drawn from the experiments before the World War I was that in contrast to a contracting muscle, the nerve pulse generates no measurable heat. Therefore, it was considered most likely that the nerve pulse consisted of a reversible physical phenomenon similar to a mechanical wave. While Hill's paper was very influential after it's publication, it is nearly forgotten today.

\begin{figure}[htbp]
	\centering
	\includegraphics[width=225pt,height=156pt]{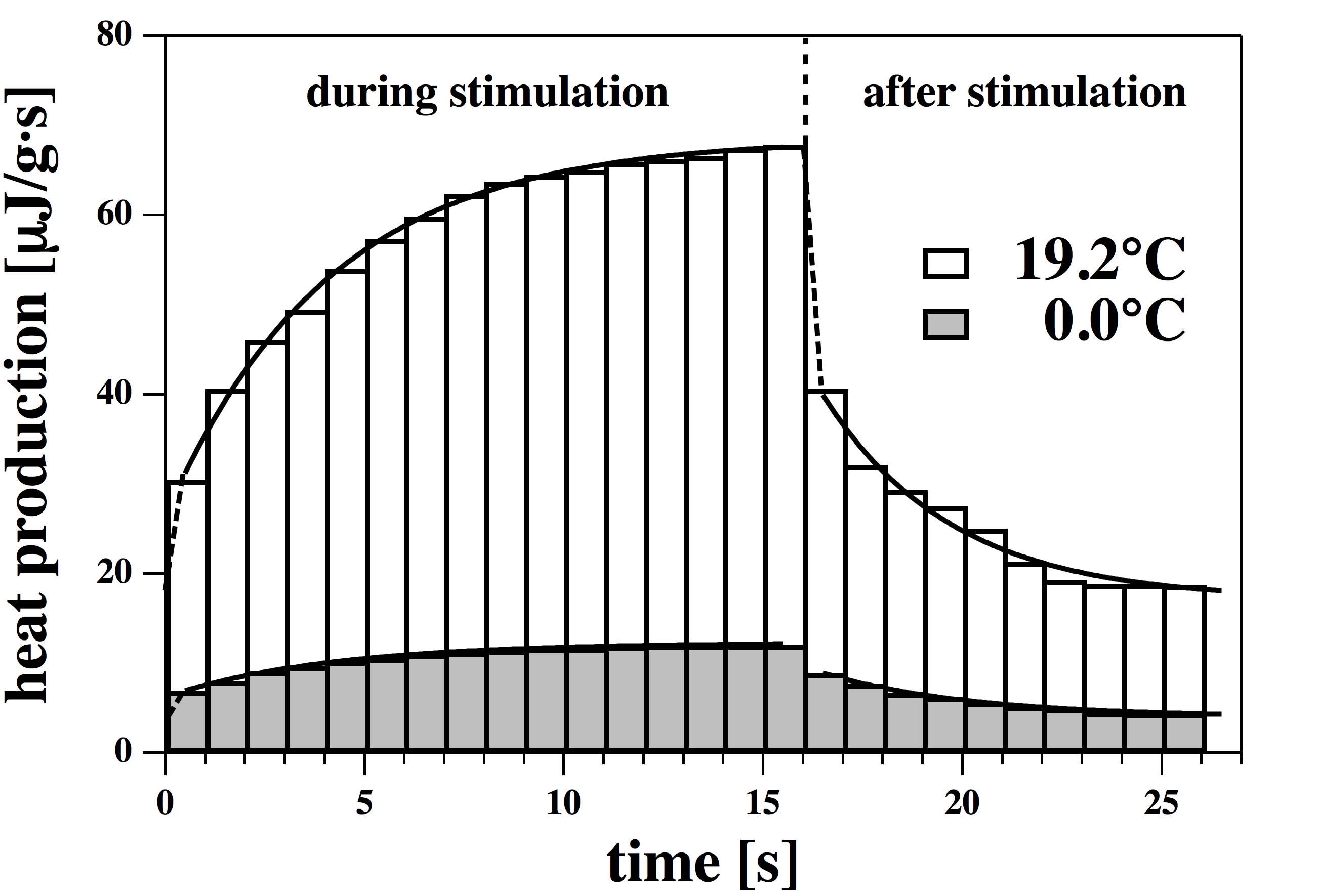}
	\caption{\small\textit {Heat production of a frog nerve during repetitive tetanus stimulation lasting 16 seconds and in the phase after the stimulation ends. The heat production of frogs adapted to 19.2$^\circ$C and 0$^\circ$ C is shown. The heat increments were measured in 1 second blocks. While there is heat production during stimulation, there is also some decaying heat production after the stimulation ended. The heat production of the frog adapted to 19.2$^\circ$C is larger then that of the frog adapted to 0$^\circ$C. Data taken from \cite[]{Hill1932}. This data do not contain the heat of the nerve pulse itself, which was too small and too fast to be measurable.}}
	\label{figure_heat1}
\end{figure}

In the following decades, Hill became the dominating authority in the measurement of heat production in muscles and nerves. In 1922, he was rewarded with the Nobel prize for physiology. The problem of the undetectable heat production of the nerve pulse continued to bother Hill. In the second half of the 1920s, he and his collaborators Downing and Gerard managed to improve the sensitivity of the temperature measurements \cite[]{Downing1926}. Between 1926 and 1935, he continued studying the problem of heat generation in nerves \cite{Downing1929, Hill1929, Hill1932, Feng1933a, Feng1933b, Feng1933c, Hill1933a, Hill1933b, Hill1934}. Since the heat production (if there should be any) was obviously very small, Hill and collaborators measured the heat production during a repetitive stimulus of the nerve. Stimulation frequencies around 70 Hz were used \cite[]{Hill1929}. A typical experiment on frog nerves is shown in Fig. \ref{figure_heat1}. There is a continuous small increase in heat production during repetitive stimulation, and there is a decaying heat production after the stimulation ceases that lasts several seconds. This suggests that the heat production is not due to the propagating pulse itself but to metabolism in the background. To make a formal distinction between the two contributions, Hill called the heat production during the impulse itself the `initial heat' and the production after the pulse (the background metabolism) the `recovery phase'. The initial heat is the heat of the propagating pulse. Since the heat data were recorded in 1 second blocks, it was not possible to resolve the heat production of the action potential itself. Hill nevertheless concluded that there is no indication for an initial heat \cite{Hill1932}:

\begin{quote}
	\emph{``The heat production of nerve is believed to occur in two phases, `initial' and `recovery'; the former is presumably an accompaniment of the physical and chemical changes which take place during the propagation of the impulse; the latter, of the processes by which those changes are reversed and the nerve restored to its initial state. It is not easy to separate the one from the other; indeed, during the earlier part of this research it was realised that in a strict sense, and on the evidence available, there might really be no ``initial'' heat at all.''}
\end{quote}

Thus, by the end of this period the heat production of the nerve pulse itself could still not be resolved and it was questionable whether it existed at all. There was, however, clear evidence of metabolism in the background related to repetitive stimulation on a time scale significantly longer than that of the nerve pulse.

\begin{figure}[htbp]
	\centering
	\includegraphics[width=225pt]{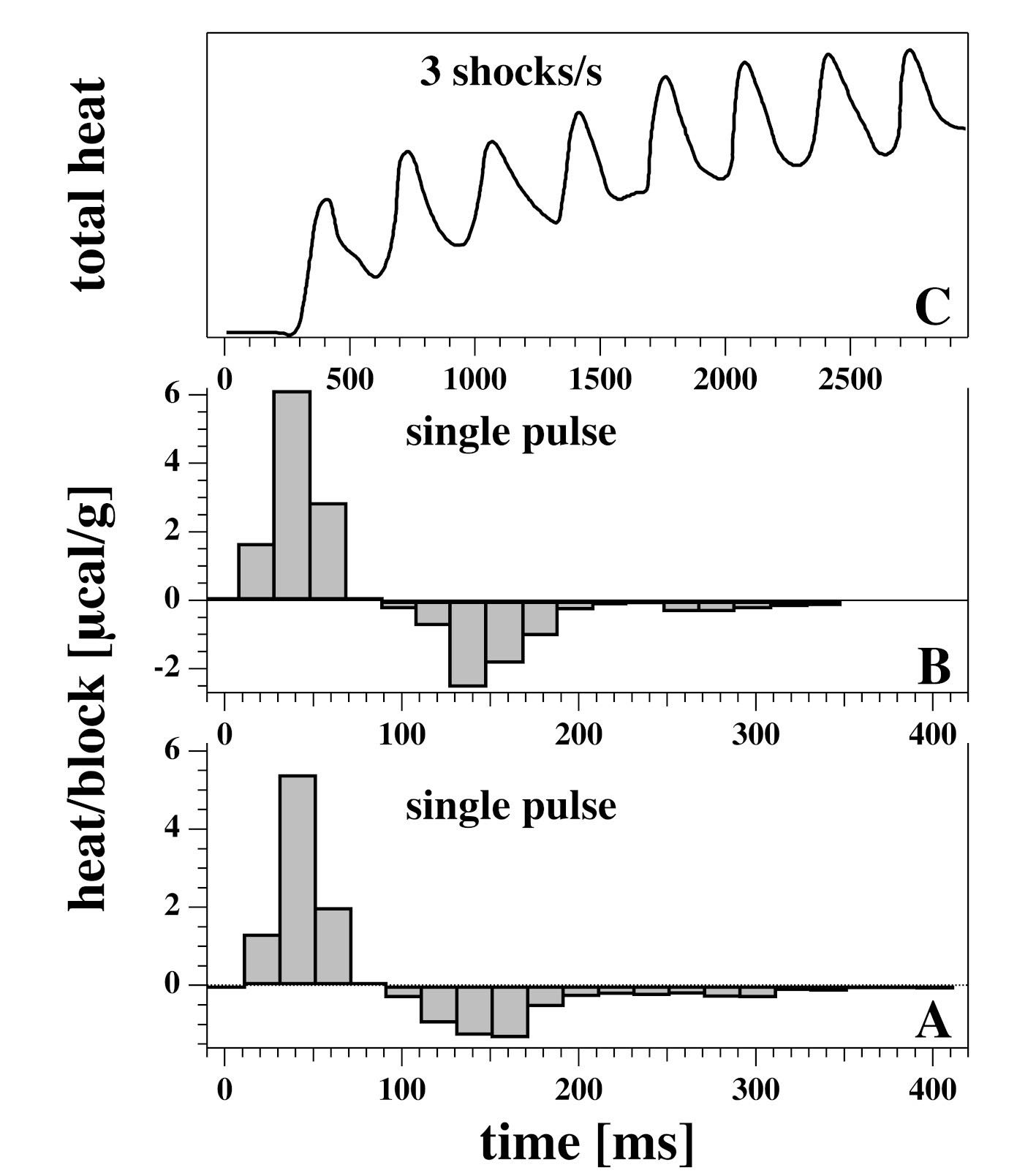}
	\caption{\small\textit {The initial heat in nerves from spider crab (adapted from \cite{Abbott1958}) Both phases can be clearly distinguished. \textbf{A, B:} Heat block analysis for two different nerves after a single stimulation \cite{Abbott1958} (average of 5 responses). The data represent the rate of heat release per block of 20 ms. It can be seen that the initial heat consists of a phase of heat release followed by a phase of heat reabsorption. The sum over all heat blocks is the net heat released. \textbf{C} Total heat release (the integral of traces as in A and B) during repetitive stimulation with 3 shocks\slash second. Each maximum corresponds to heat release and reabsorption of heat. The slow drift represents the difference of heat release and absorption after a pulse (shown in Fig.\,\ref{figure_heat1}), which might be due to metabolism.}}
	\label{abbott1958}
\end{figure}

In a seminal publication in 1958, Abbott, Hill and Ho-\linebreak warth \cite{Abbott1958} (see also \cite{Hill1958}) first resolved the initial heat response during the action potential in nerves from the legs of spider crabs. They called their article ``The positive and negative heat production associated with a nerve impulse''. It was found that, during the rising phase of the action potential, heat is released into the environment of the nerve. This heat is mostly reabsorbed (to at least 70 \%) in a second phase of the action potential. Some experiments from \cite{Abbott1958} are shown in Fig. \ref{abbott1958}. The existence of a reabsorption of heat is clearly surprising.

At the time it was not clear whether the heat is completely reabsorbed, or whether some heat is dissipated. Further, there existed no satisfactory explanation for both the qualitative and quantitative behavior of the heat production.

Hodgkin acknowledged the problem arising from Abbott et al.'s result in his monograph ``the nervous impulse'' from 1964 \cite[p.70]{Hodgkin1964}

\begin{quote}
	\emph{``In thinking about the physical basis of the action potential perhaps the most important thing to do at the present moment is to consider whether there are any unexplained observations which have been neglected in an attempt to make the experiments fit into a tidy pattern. {\ldots} perhaps the most puzzling observation is one made by A. V. Hill and his collaborators Abbott and Howarth (1958). {\ldots} On reinvestigating the initial heat of crab nerve with better time resolution, Hill and his colleagues found that it was diphasic and that an initial phase of heat liberation was followed by one of heat absorption. {\ldots} a net cooling on open-circuit was totally unexpected and has so far received no satisfactory explanation.''}
\end{quote}

\begin{figure}[htbp]
	\centering
	\includegraphics[width=225pt]{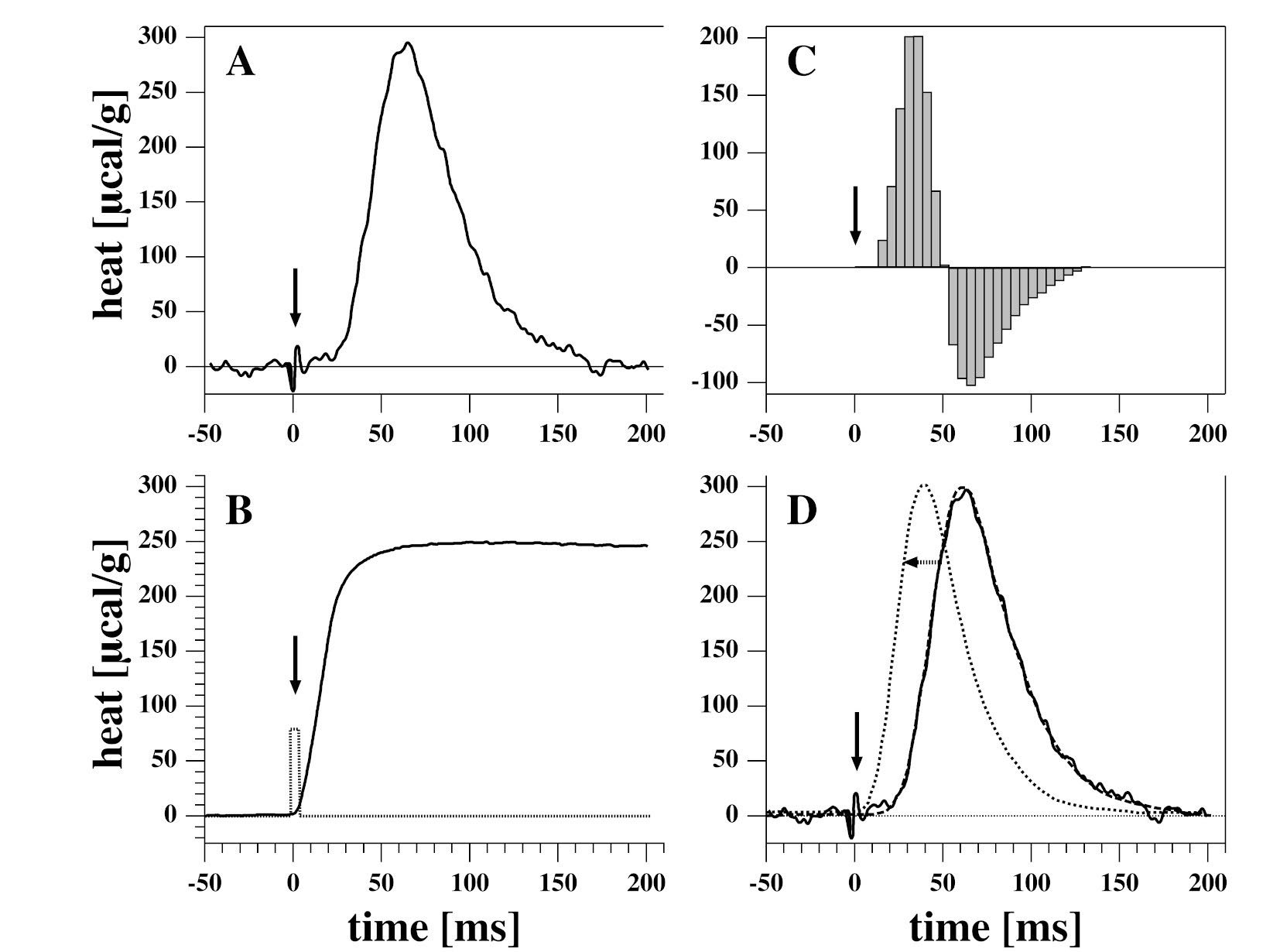}
	\caption{\small\textit {Heat production in non-myelinated fibers of garfish olfactory nerve (adapted from \cite{Ritchie1985}). (A) Thermal response after a stimulation indicated by the stimulation artifact. (B) Control experiment after applying a local heated pulse on the membrane. (C) Heat flux out and into the nerves obtained by heat block analysis. (D) Deconvolution of the measured signal using the response function of the instrument leading to the true signal (dashed line). The derivative of this signal corresponds to the true heat flux in (C). }}
	\label{heat_deconvolution_ritchie1985}
\end{figure}

During the next 30 years, Hill's studies were continued and were refined by close collaborators, in particular by \linebreak Howarth, Abbott and Ritchie \cite{Abbott1965, Howarth1968, Abbott1973, Ritchie1973, Howarth1975, Howarth1975b, Howarth1979a, Howarth1979b}. Temperature changes were recorded by a series of many (i.e. 96 \cite{Abbott1958}) thermocouples. The thermocouples were insulated from electrical contact by embedding them into a raisin. Since raisin has bad heat conduction features and the thermocouples have a heat capacity of their own, one has to consider the distortion of the recordings by the setup itself. In order to obtain the true heat change, the results of the measurement have to be deconvoluted using the response function of the instrument (Fig. \ref{heat_deconvolution_ritchie1985}). Therefore, all of these studies used a so-called heat block analysis (shown in Fig.\,\ref{abbott1958}), which is a kind of empirical deconvolution of the measured data and the response function of the instrumentation (Fig. \ref{heat_deconvolution_ritchie1985}, B). It is described in detail in \cite{Howarth1975}. Due to the long response time of the setup ($\ge 60$ ms), these experiments are limited to nerves with very long action potentials such as the rabbit vagus nerve or garfish olfactory nerve at low temperature with pulse widths of about 100--200 ms.

In all of the above papers the authors came to the conclusion that the charging and discharging of a membrane capacitor (expected in the electrical models for nerve conduction \cite{Hodgkin1952b}) cannot explain the observed changes in heat. The authors propose that most of the reversible heat production is due to entropy changes within the membrane during the action potential (see section \ref{thecondensertheory}).

\begin{figure}[htbp]
	\centering
	\includegraphics[width=225pt,height=172pt]{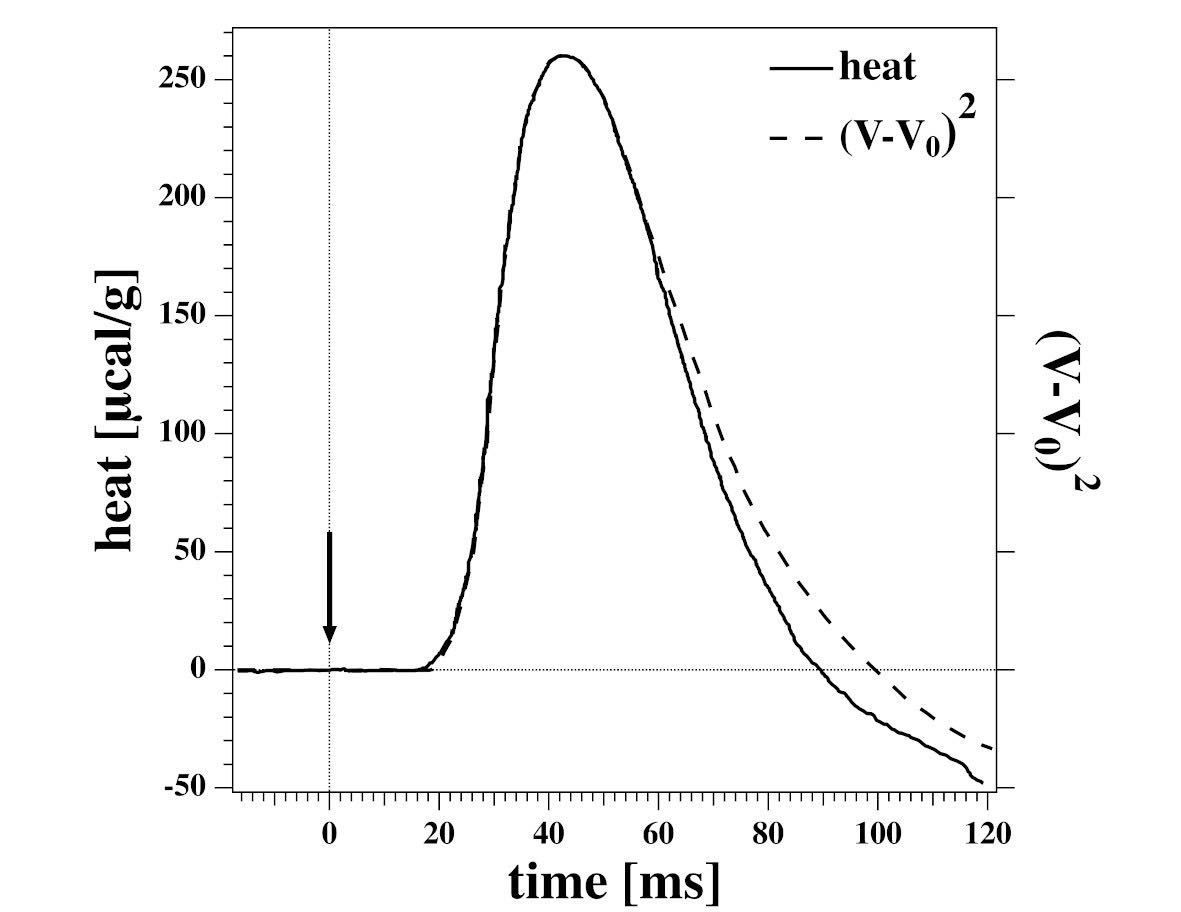}
	\caption{\small\textit {Deconvoluted heat response of a garfish olfactory nerve (solid line) compared to the voltage of the pulses ($V^2-V_0^2$) in arbitrary units. The heat signal is in phase with the voltage changes and proportional to the energy of the charged capacitor. Adapted from \cite{Ritchie1985}. }}
	\label{heat_voltage_ritchie1985}
\end{figure}

The essential findings of the heat experiments are summarized in a review by Ritchie \& Keynes from 1985 \cite{Ritchie1985}. They ruled out the possibility that the heat reabsorption phase of the action potential was due to heat diffusion into the bulk medium. The local heating of a nerve showed that heat diffusion away from the nerve is much slower than the reabsorption of heat that is measured during the action potential. This control experiment is shown in Fig. \ref{heat_deconvolution_ritchie1985} (B) \cite{Ritchie1985}. They also found that the action potential is exactly in phase and approximately proportional to the square of the transmembrane voltage (Fig. \ref{heat_voltage_ritchie1985}). Thus, the heat release is proportional to the energy of charging and discharging a capacitor. After some initial doubts in earlier papers on whether the heat reabsorption is complete, they also found that all heat is reabsorbed within experimental accuracy. In fact, they could show that, even for action potentials with a hyperpolarization phase, the total heat followed voltage (Fig. \ref{heat_voltage_ritchie1985}) \cite{Ritchie1985}.

\begin{figure*}[htbp]
	\centering
	\includegraphics[width=394pt,height=120pt]{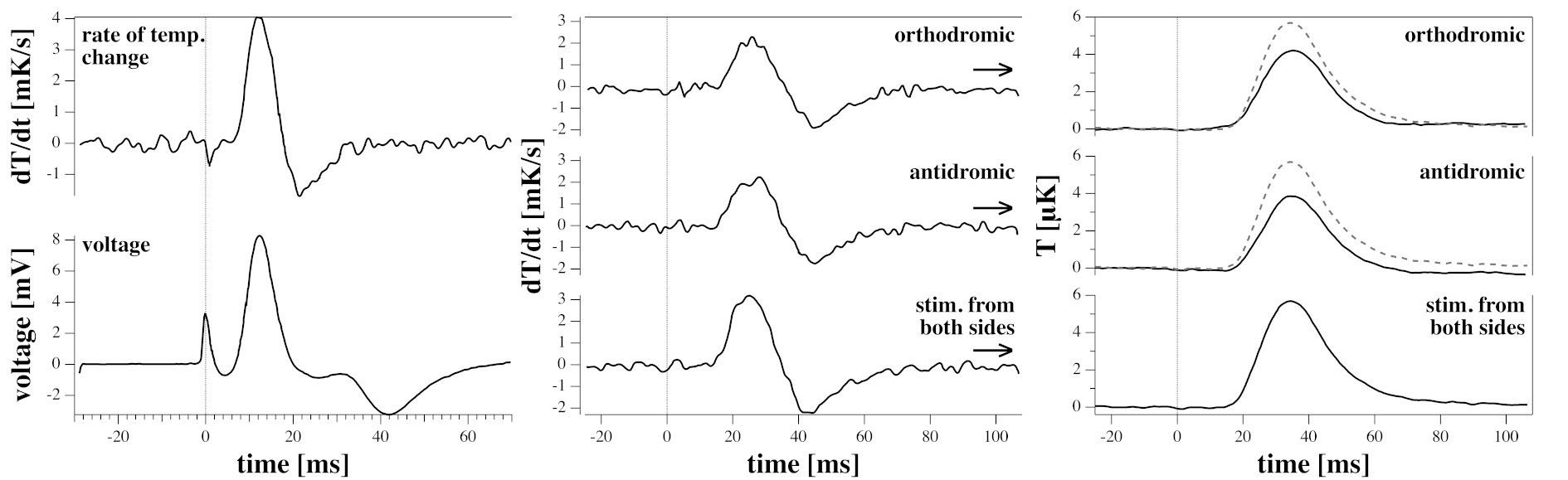}
	\caption{\small\textit {Thermal response of garfish olfactory nerve (adapted from \cite{Tasaki1989}). Stimulation occurs at a time of 0 ms. Left: rate of temperature change at 20$^\circ$C (trace a) and the accompanying action potential. The temperature change and the voltage are nearly in phase. Center: rate of temperature change for an orthodromic (A), antidromic (B) and combined (C) nerve propagation experiment at 10$^\circ$C. Right: the integrated rate yields the absolute temperature change, which is of the order of 40 \textmu K and nearly completely reversible. The stimulation from both sides yields a larger temperature change than stimulation from one end. This is indicated by the grey dashed line in the top and middle panel on the right.}}
	\label{tasaki1989}
\end{figure*}

The above experiments were limited to long action potentials of 100--200 ms. However, most nerve impulses display a temporal length on the order of 1 ms, which is about 2 orders of magnitude faster. Measurement of the heat release in such nerves requires different methods. Tasaki and Iwasa managed to obtain a faster response time by using a heat-sensitive polyvinylidene fluoride film \cite{Tasaki1981}. They recorded an action potential of 25 ms length in lobster nerve and later in garfish olfactory nerve and myelinated nerve fibers from bullfrog on action potentials with a time scale of 10--20 ms \cite{Tasaki1989}. As in the earlier measurements by Abbott, Howarth and Ritchie, Tasaki and collaborators found that the integrated heat is very small and consistent with nearly complete reabsorption. Results are shown in Fig.\,\ref{tasaki1989}. These authors concluded that the heat production is in phase with the nerve pulse but that the total heat is unlikely to be caused by charging and discharging the membrane capacitor. The time course of voltage and heat was not always identical (Fig.\,\ref{tasaki1989} left a and b). They proposed that the heat production might originate from ion exchange in the superficial layer of the axoplasm \cite{Tasaki1989} which was an idea generally favored by Tasaki. He proposed that the nerve pulse is accompanied by a phase transition in the proteins on the surface of the nerve \cite{Tasaki1999}.

\subsection{Other important thermodynamic findings on nerves}
\label{otherimportantthermodynamicfindingsonnerves}

It should be noted that, in addition to the striking reabsorption of heat, various other experiments indicate that the nerve pulse is not a purely electrical phenomenon. It has been found that the membrane of the axon changes its thickness \cite{Iwasa1980a, Iwasa1980b, Tasaki1982d, Tasaki1982e, Tasaki1982b,Tasaki1989, GonzalezPerez2016} and its length \cite{Wilke1912b, Tasaki1980, Tasaki1989}, i.e., a change in axon dimension. Further, it has been shown in birefringence, light scattering and fluorescence experiments that lipids in the membrane change orientation and order during the pulse \cite{Cohen1968, Tasaki1968, Tasaki1969a, Tasaki1969b,  Tasaki1971, Cohen1971}. This indicates that the physical reality in nerves is far richer than that assumed in a purely electrical picture.


\section{Models for the nervous impulse}
\label{modelsforthenervousimpulse}

There exist two major theories for the function of nerves. The first is the Hodgkin-Huxley model \cite{Hodgkin1952b}. It is a molecular theory that employs ion channel proteins acting as transistors and semiconductors. The Hodgkin-Huxley model is a purely electrical theory that does not explicitly mention heat, temperature or changes in spatial dimensions.

The second theory is the soliton model that considers the action potential as an adiabatic sound pulse during which all thermodynamic variables change in a manner that conserves entropy \cite{Heimburg2005c, Lautrup2011}. This is consistent with the view that Hill expressed in 1912 \cite{Hill1912}. It is based on the existence of a phase transition in the axonal membrane. The heat production is related to the latent heat of this transition.


\subsection{The Hodgkin-Huxley model}
\label{thehodgkin-huxleymodel}

\begin{figure}[htbp]
	\centering
	\includegraphics[width=169pt,height=120pt]{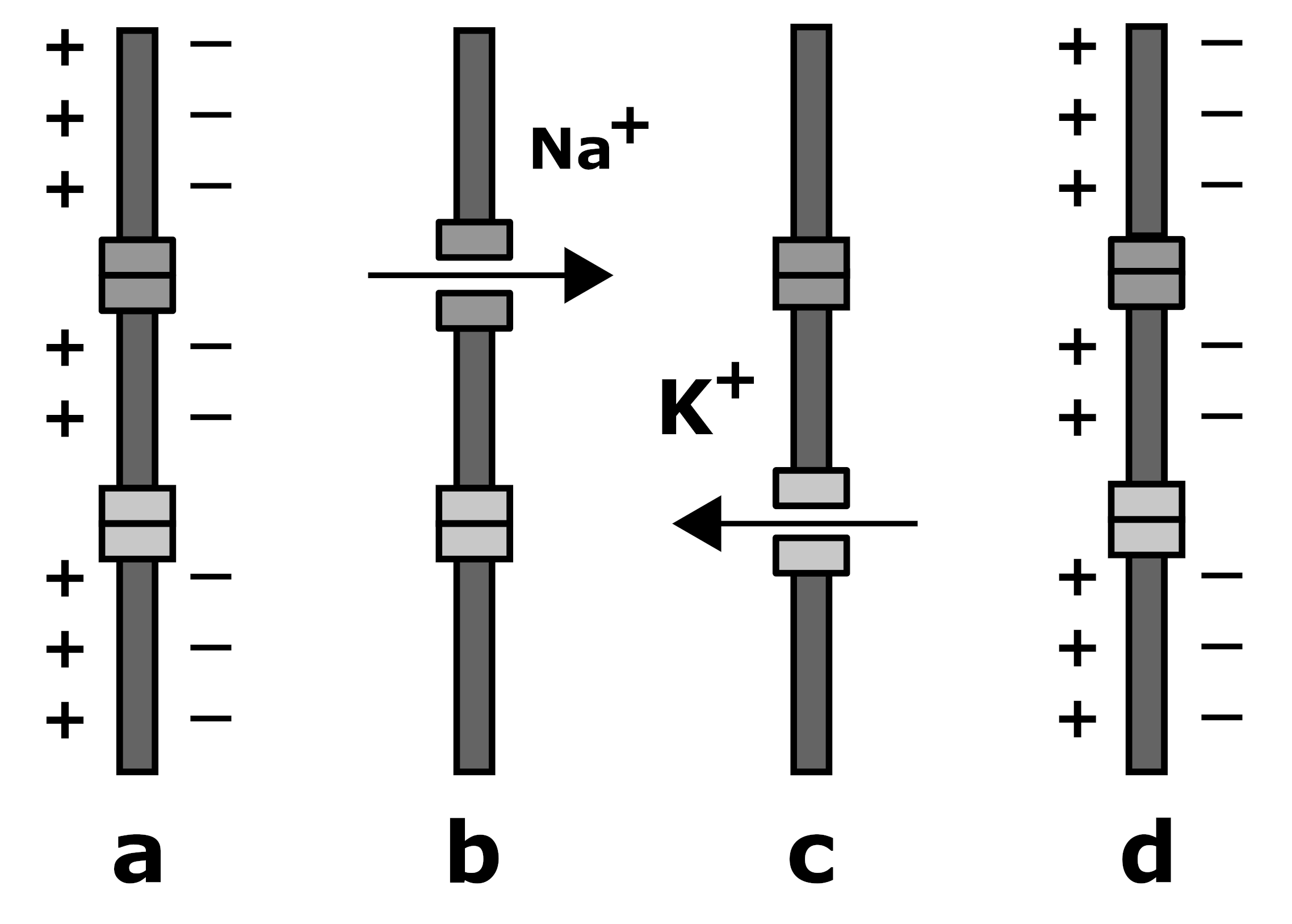}
	\caption{\small\textit {Sequence of events in the Hodgkin-Huxley model. \textbf{a.} Charged axon membrane with a resting potential V$_R$ and two channels for sodium and potassium ions which are both closed. \textbf{b.} Sodium flows into the axon while the sodium channels open. The membrane capacitor is discharged. \textbf{c.} Sodium channels close and potassium channels open. Potassium flows from the inside to the outside while the membrane is recharged. \textbf{d.} The membrane is back in its resting state.}}
	\label{zeichnung}
\end{figure}

The Hodgkin-Huxley model \cite{Hodgkin1952b} presently is the most commonly accepted model for the action potential. It is purely electrical and based on so-called equivalent circuits which replace biological objects by electrical analogies. It contains voltage-gated resistors called ion channel proteins (semi-con\-ductors), a capacitor (the nerve membrane) and batteries (the difference in the concentration of sodium and potassium ions between the inside and the outside of the axon). The action potential is generated by the voltage-dependent opening and closing of the channels after electrical stimulation. The resulting fluxes of sodium and potassium across the membrane briefly change the charge on the membrane capacitor and hence the transmembrane voltage. Fig. \ref{zeichnung}. shows a discharging (depolarization) via the opening of sodium channels and recharging of the axonal membrane (repolarization) via the opening of potassium channels. This together with cable theory \cite{Johnston1995} leads to a propagating voltage pulse.

The Hodgkin-Huxley model does not include any effects of temperature changes. When considering the heat response in such a system, one would have to consider heat generation by currents through resistors caused by friction, the temperature changes in the batteries and the energy of the capacitor --- i.e., one would have to translate the equivalent circuits into their thermodynamic analogs. The batteries would then correspond to the expansion of sodium and potassium as ideal gases across semipermeable walls and the charging of a capacitor as the corresponding work. This is discussed in section \ref{thecondensertheory}.


\subsection{The soliton model}
\label{thesolitonmodel}

\begin{figure}[htbp]
	\centering
	\includegraphics[width=225pt,height=62pt]{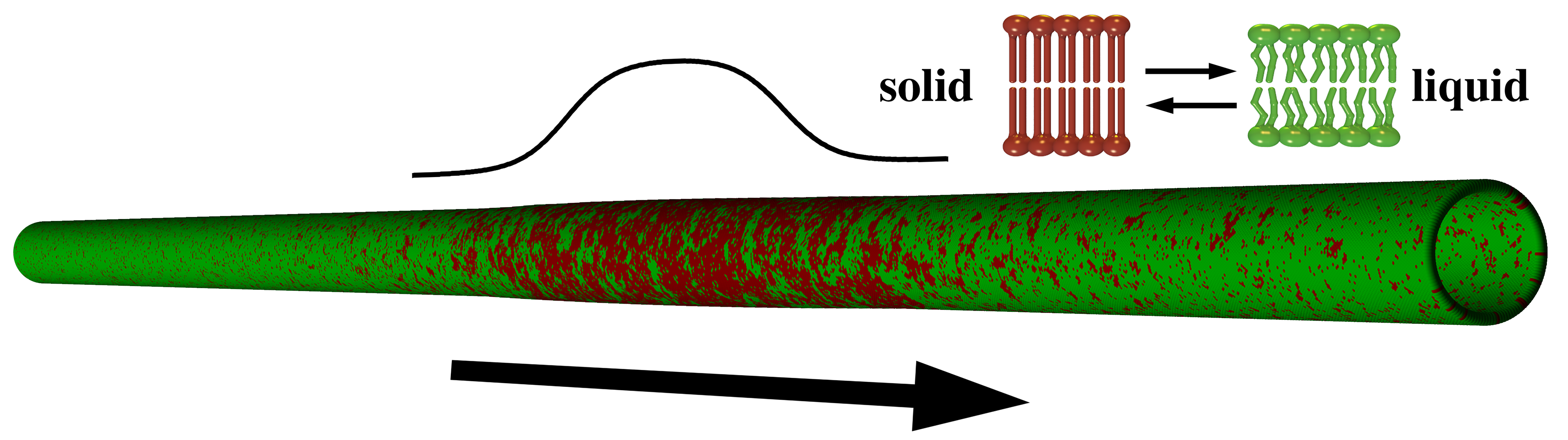}
	\caption{\small\textit {Schematic representation of a propagating soliton in a membrane cylinder. The pulse consists of a solid lipid region traveling in a liquid membrane environment. The black line is a schematical representation of the changes in membrane density (see also Fig. \ref{soliton_heat}). From \cite{GonzalezPerez2016}.}}
	\label{soliton}
\end{figure}

The soliton model \cite{Heimburg2005c, Lautrup2011} is based on a thermodynamic theory that includes both thermal, mechanical and electrical effects. At physiological temperatures, biological membranes exist in a fluid state slightly above a solid-liquid transition \cite{Muzic2019}. The soliton model assumes that the nerve pulse consist of ordered (gel) region of lipids traveling in a disordered (fluid) membrane environment (see Fig. \ref{soliton}), reminiscing of density waves and sound. Since the membrane becomes ordered in the region of the pulse, the latent heat of the transition is released. When the pulse has passed, the membrane returns to its liquid state and reabsorbs the latent heat in a fully reversible manner without dissipation of heat. Since the solid membrane is thicker and has as smaller area than the liquid membrane, the nerve contracts during the action potential and its thickness increases. The ordered regions of the soliton have different lipid anisotropies that are visible in fluorescence experiments. Electrical properties of the nerve membrane emerge due to changes in capacitance which are a consequence of thickness and area changes and changes in electrical polarization when moving from fluid to solid phase. One finds a 50\% reduction in capacitance \cite{Zecchi2017} . Capacitance and polarization changes can alter the voltage without net flow of charge through the membrane.

\begin{figure}[htbp]
	\centering
	\includegraphics[width=225pt,height=183pt]{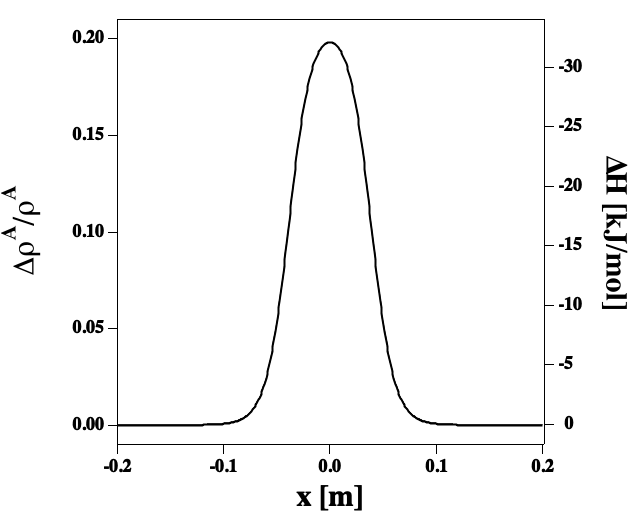}
	\caption{\small\textit {The change in area density of an electromechanical soliton in comparison to the magnitude of the heat released into the environment per mol of lipid.}}
	\label{soliton_heat}
\end{figure}

The soliton model describes the action potential as an adiabatic and isentropic pulse. This implies that heat is not dissipated but there is a reversible conversion of work into heat. In section \ref{heatreleaseinthesolitonmodel} we will investigate the magnitude of heat changes during the action potential that are associated with the soliton model. The density profile and the heat release are shown in Fig.\,\ref{soliton_heat}. The soliton model is qualitatively consistent with all of the above measurements.


\section{The thermodynamics of adiabatic \\processes}
\label{thethermodynamicsofadiabaticprocesses}

In the following we will introduce some basic concepts for the thermodynamics of compressible systems with a special focus on adiabaticity. We will start with adiabatic compression of gases, then consider adiabatic compression of layers and springs, and discuss the reversible charging of capacitors. These concepts will then be applied to nerves.


\subsection{Compressible systems}
\label{compressiblesystems}

The first law of thermodynamics states that
\begin{equation}
dE=TdS - pdV - \Pi dA - Fdx + \Psi dq + ... \;,
\label{eq:adiab1}
\end{equation}

where $E$ ,$S$, $V$, $A$, $x$ and $q$ are the extensive quantities internal energy, entropy, volume, area, position and charge. $T$, $p$, $\Pi$, $F$ and $\Psi$ are the intensive variables temperature, pressure, lateral pressure, force, and electrostatic potential. The differential of the internal energy, $dE=TdS-pdV$ (constant area, charge, etc), is given by
\begin{eqnarray}
dE&=&T\left(\frac{\partial S}{\partial T}\right)_VdT+T\left(\frac{\partial S}{\partial V}\right)_TdV-pdV\nonumber\\
&=&c_VdT+\left[T\left(\frac{\partial p}{\partial T}\right)_V-p\right]dV
\label{eq:adiab1.1}
\end{eqnarray}
where we made use of the heat capacity at constant volume, $c_V=T(\partial S/\partial T)_V$, and the Maxwell relation $(\partial S/\partial V)_T=(\partial p/\partial T)_V$ \cite{Sommerfeld1992e}. When the compression is adiabatic, $dE=-pdV$. The pressure is generally temperature-dependent. The sum of the first two terms on the right hand side of eq. (\ref{eq:adiab1.1}) are zero (because $dS=0$), which provides a condition for the relation between temperature and volume changes. Under adiabatic compression, the temperature will change provided that $\left(\partial p/\partial T\right)_V$ is non-zero. At constant volume $V$,
\begin{eqnarray}
dV &=& \left(\frac{\partial V}{\partial T}\right)_p dT +\left(\frac{\partial V}{\partial p}\right)_T dp=0\nonumber\\
&&\mbox{so that}\nonumber\\
\left(\frac{\partial p}{\partial T}\right)_V&=&\alpha_p K_T^V
\label{eq:adiab1.2}
\end{eqnarray}
where $\alpha_p=(\partial V/\partial T)_p$ is the thermal expansion coefficient at constant pressure and $K_T^V= -(\partial p/\partial V)_T$ is the isothermal compression modulus. Both $\alpha_p$ and $K_T^V$ are different from zero for all materials. Therefore, the temperature change due to adiabatic compression is always different from zero.

In the case of an adiabatic compression,
\begin{eqnarray}
dS&=&\frac{c_V}{T} dT + \left(\frac{\partial p}{\partial T}\right)_V dV=0\qquad\nonumber\\&\mbox{so that}&\nonumber\\
dT &=& -\frac{T}{c_V}\left(\frac{\partial p}{\partial T}\right)_V dV
\label{eq:adiab1.3}
\end{eqnarray}


\subsubsection{Gases}
\label{gases}

\begin{figure}[htbp]
	\centering
	\includegraphics[width=225pt,height=114pt]{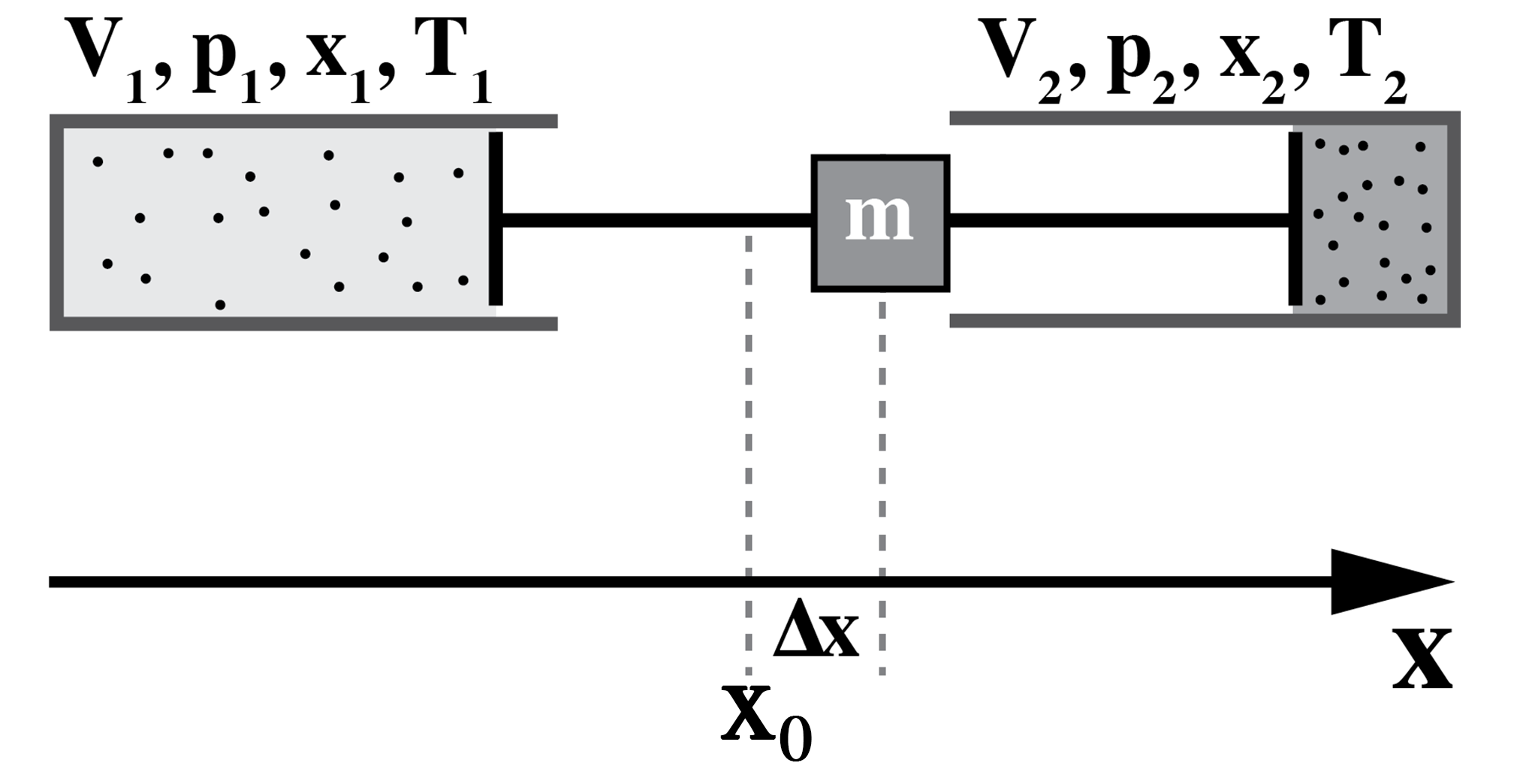}
	\caption{\small\textit {Two gas containers connected to a moving mass $m$. }}
	\label{gaspiston}
\end{figure}

The equation of state for an ideal gas is
\begin{equation}
pV=NkT
\label{eq:ideal1}
\end{equation}
from which follows that $(\partial p/\partial T)_V=N k/V$. The heat capacity of an ideal monoatomic gas is simply $c_V=3/2 N k$. During the adiabatic compression of a gas from volume $V_0$ to $V$, the temperature changes from $T_0$ to $T$. From eq. (\ref{eq:adiab1.3}) we find that

\begin{equation}
T^{3/2}\cdot V =T_0^{3/2}\cdot V_0 = {\rm const}
\label{eq:ideal4}
\end{equation}

\noindent This is the well-known equation of state for the adiabatic compression of an ideal gas. It states that the work done on an ideal gas when it is compressed is converted into thermal energy of the gas particles.

\begin{figure}[htbp]
	\centering
	\includegraphics[width=225pt,height=175pt]{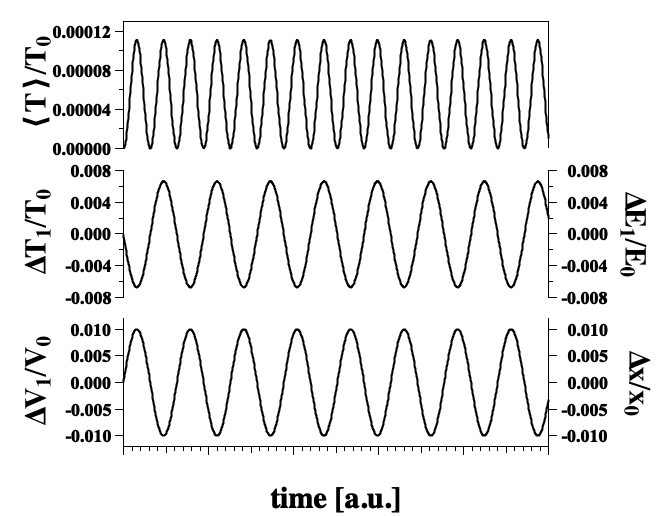}
	\caption{\small\textit {The temperature in the two gas containers in Fig. \ref{gaspiston} (left) and the energy of both containers combined (top trace). Adapted from \cite{Heimburg2017}.}}
	\label{gaspistontemp}
\end{figure}

One can use the above to determine the change in temperature in two coupled gas pistons that undergo oscillations (see Fig. \ref{gaspiston}). In this system one can determine either the temperature in each container ($T_1$ and $T_2$) or the average temperature change of both containers, $\left<T\right>=(T_1+T_2)/2$. One obtains temperature variations of the gas containers as shown in Fig. \ref{gaspistontemp}. \\

Similar relations can be found for real gases. For instance, the equation of state for a van der Waals-gas is given by
\begin{equation}
\left(p+\frac{N^2 a}{V^2}\right)(V-Nb)=NkT
\label{eq:VdW1}
\end{equation}
where $b$ is the excluded volume per gas particle, $N$ is the number of particles and $a$ represents the effects of an attractive or repulsive force. Using eq.(\ref{eq:adiab1.3}) and the fact that $c_V=3/2 Nk$ for the vdW gas, we obtain
\begin{equation}
\frac{T}{T_0}=\left(\frac{V_0-Nb}{V-Nb}\right)^{2/3}
\label{eq:VdW4}
\end{equation}
Interestingly, the temperature changes do not depend on the forces between particles. They depend only on the volume changes. We see that adiabatic oscillation of systems involving both ideal and real gases will show periodic temperature variations in the gas volume.


\subsubsection{Membranes}
\label{membranes}

Using the same strategy for two-dimensional systems such as membranes, the differential of the internal energy change is given by $dE=TdS-\Pi dA$ and
\begin{equation}
dE=c_A dT+\left[T\left(\frac{\partial \Pi}{\partial T}\right)_A-\Pi\right]dA \;,
\label{eq:adiabmem1}
\end{equation}
where $\Pi$ is the lateral pressure, $A$ is the area of the layer and $c_A$ is the heat capacity at constant area. In the special case of biological membranes one has to consider that they are surrounded by water. Therefore, the heat capacity of the interfacial water is part of $c_A$ \cite{Mosgaard2013a}.

The change in temperature upon adiabatic lateral compression is given by
\begin{equation}
dT = -\frac{T}{c_A}\left(\frac{\partial \Pi}{\partial T}\right)_A dA
\label{eq:adiabmem3}
\end{equation}
with
\begin{equation}
\left(\frac{\partial \Pi}{\partial T}\right)_A=\alpha_{\Pi} K_T^A
\label{eq:adiabmem4}
\end{equation}
where $\alpha_{\Pi}=(\partial A/\partial T)_{\Pi}$ is the thermal area expansion coefficient at constant lateral pressure and $K_T^A= -(\partial \Pi/\partial A)_T$ is the isothermal lateral compression modulus.

Eqs. (\ref{eq:adiabmem1}) and (\ref{eq:adiabmem3}) indicate that the extensive variable controlled in the experiment is the area. However, the pressure is a function that depends on all other variables. For instance, a charged membrane displays a different lateral pressure as compared to an uncharged membrane with the same area. If area is controlled, the other variables (which are not controlled) will adapt to maximize entropy. This is also true for all previous and subsequent arguments. The fact that one variable is controlled from the outside does not imply that the other variables are fixed and constant under compression. In this respect, it is not meaningful to regard mechanical changes as being separate from electrical changes. The compressibility and the area expansion coefficient in Eq. (\ref{eq:adiabmem4}) both contain the electrostatics (and also the chemistry) of the membrane. This is also true for the soliton model, in which variations of the different variables are all contained in pressure and compressibility.

Materials increase their temperature upon compression. Simultaneously, material expand upon heating. For this reason, $(\partial S/\partial A)_T=(\partial \Pi/\partial T)_A$ is related to both area expansion coefficient $\alpha$ and the compressibility $\kappa_T^A$. In order to be compressible, the compressed material needs a heat sink with a finite heat capacity. Otherwise, the increase in temperature would lead to an expansion of the membrane. The case of membranes is therefore particularly interesting because it is a two-dimensional system embedded into a three-dimensional medium. The medium can serve as a heat buffer that contributes to the compression modulus of the membrane. This has been discussed in detail in \cite{Mosgaard2015a}.

Adiabatic waves in a membrane (to be discussed below) will cause temperature variations with a magnitude related to the thermal area expansion coefficient and the isothermal area compression modulus. Therefore, the compression of a membrane will lead to temperature changes in its environment.


\subsubsection{Springs}
\label{springs}

\begin{figure}[htbp]
	\centering
	\includegraphics[width=84pt,height=184pt]{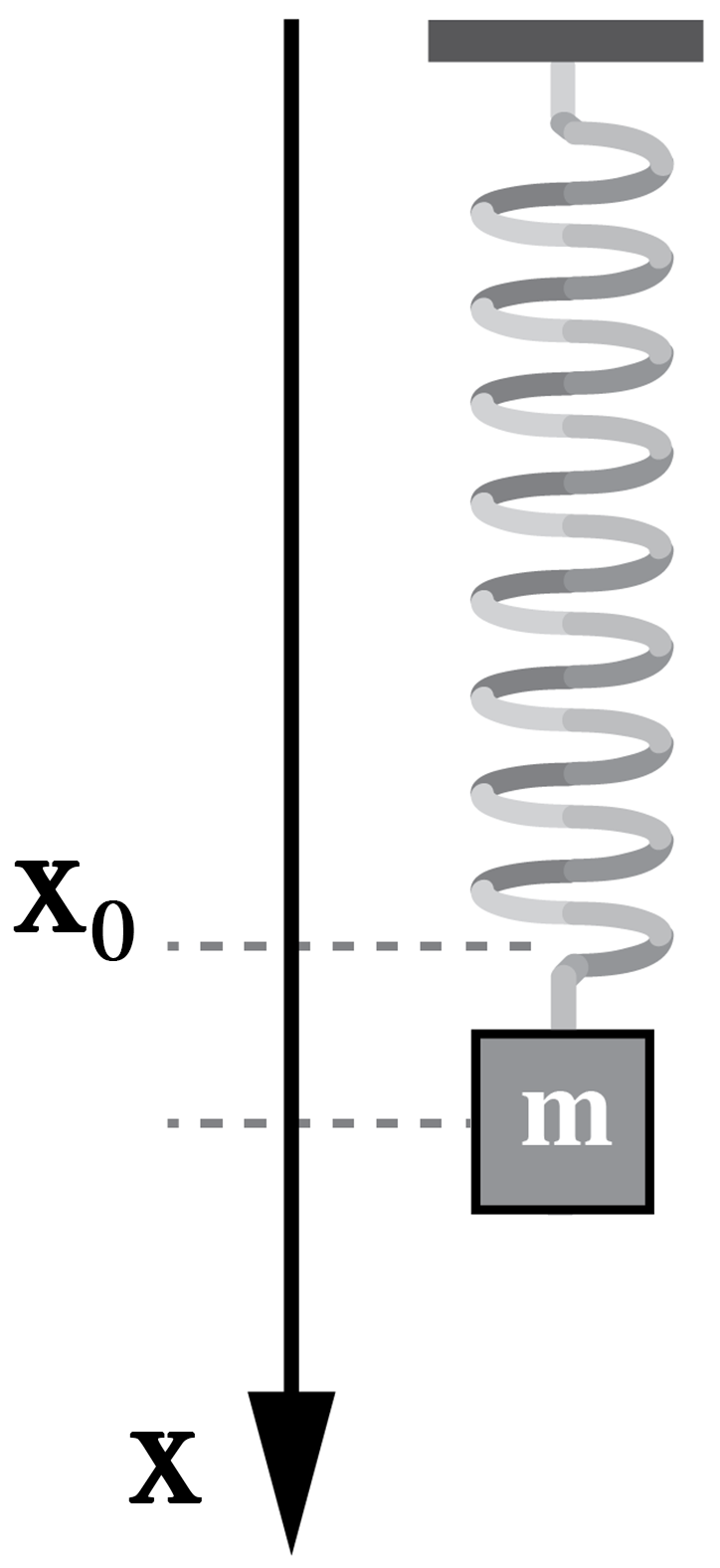}
	\caption{\small\textit {Mass connected to a spring}}
	\label{springmass}
\end{figure}

Similarly, for the compression of a spring
\begin{equation}
dE=c_x dT+\left[T\left(\frac{\partial F}{\partial T}\right)_x-F\right]dx \;,
\label{eq:adiabspring1}
\end{equation}
where $c_x$ is the heat capacity at constant spring extension and $F$ is the force in the spring. In the case of adiabatic compression yields a temperature change of
\begin{equation}
dT = -\frac{T}{c_x}\left(\frac{\partial F}{\partial T}\right)_x dx
\label{eq:adiabspring2}
\end{equation}
introducing the heat capacity at constant spring extension, $c_x$. We find that
\begin{equation}
\left(\frac{\partial F}{\partial T}\right)_x=\alpha_F K_T^x \;,
\label{eq:adiabspring3}
\end{equation}
where $\alpha_F=(\partial x/\partial T)_F$ is the linear thermal expansion coefficient of a spring at constant force, and $K_T^x= (\partial F/\partial x)_T$ is the isothermal compression modulus of the spring, i.e., the isothermal spring constant. For constant entropy, $dE=-Fdx$, i.e., the internal energy of the spring is equal to the work necessary to compress it.

\begin{figure}[htbp]
	\centering
	\includegraphics[width=169pt,height=159pt]{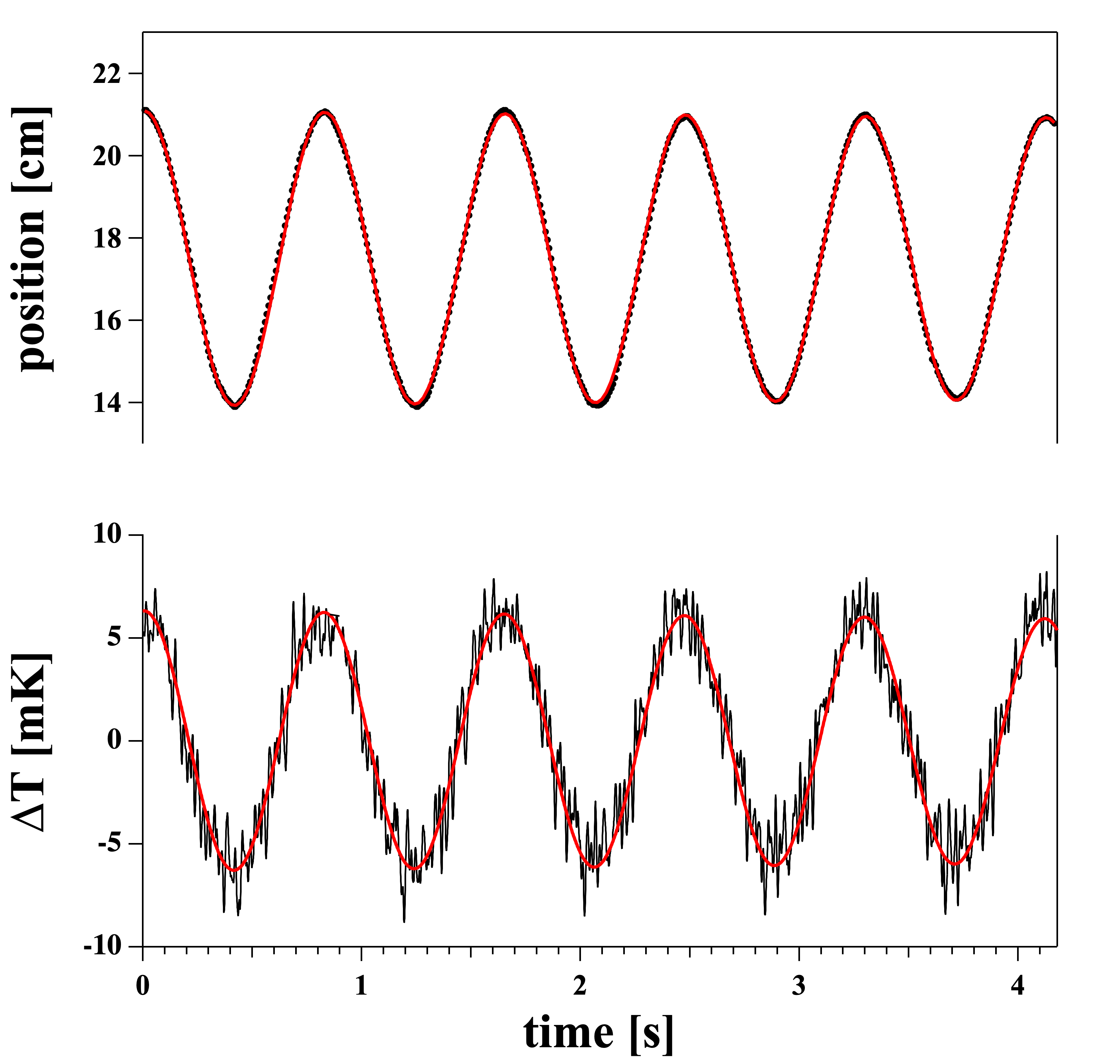}
	\caption{\small\textit {Top: Position changes of the spring described in the text. Bottom: Associated temperature changes as measured with a thermocouple. The red line in the lower panel is a sinusoidal fit to the temperature data.}}
	\label{heat_spring}
\end{figure}

The temperature change of a spring connected to a mass (Fig. \ref{springmass}) is shown in Fig. \ref{heat_spring}. The experiment was performed on a steel spring with an (adiabatic) spring constant of K=280 N\slash m attached to a weight of 5 kg. The spring has a heat capacity of 40 J\slash K. Temperature changes were measured with a tiny K-type thermocouple glued to the spring with a thermally conducting but electrically insulating film with a thickness of about 50 \textmu m (such as used for establishing heat contact of computer chips with a heat sink). Since the film has finite heat conductivity, the signal shown in Fig. \ref{heat_spring} is a lower estimate of the true temperature change.

In the case of an ideal gas considered in section \ref{gases}, compression by the piston imparts additional kinetic energy to the gas molecules. In the present case, the deformation of a spring from its equilibrium configuration imparts energy to the normal modes of the underlying crystal lattice (i.e., creates phonons). Since this effect is several orders of magnitude smaller than the total energy of the system, it is usually ignored in the elementary descriptions familiar introductory courses in classical mechanics. It is nonetheless present and can be measured as shown in Fig. \ref{heat_spring}.


\subsection{Electrical systems: Adiabatic charging of a capacitor}
\label{electricalsystems:adiabaticchargingofacapacitor}

For a system in which the variables are temperature and charge (e.g., the charging of a capacitor such as in nerves), $dE=TdS+\Psi dq$ (where $\Psi$ is the electrical potential and $q$ is a charge), and
\begin{equation}
dE=c_q dT+\left[-T\left(\frac{\partial \Psi}{\partial T}\right)_q+\Psi\right]dq
\label{eq:adiab2.1}
\end{equation}
where we define the heat capacity of the capacitor at constant charge as $c_q=T(\partial S/\partial T)_q$ and made use of the Maxwell relation $(\partial S/\partial q)_T=-(\partial \Psi/\partial T)_q$.
In the case of an adiabatic charging of a capacitor with capacitance $C_m$,
\begin{eqnarray}
dS&=&\frac{c_q}{T} dT - \left(\frac{\partial \Psi}{\partial T}\right)_q dq=0\nonumber\\
&&\mbox{so that}\nonumber\\
dT &=& -\frac{T}{c_q}\left(\frac{\partial \Psi}{\partial T}\right)_q dq \;.
\label{eq:adiab2.2}
\end{eqnarray}
The temperature dependence of the potential can be written as
\begin{equation}
\left(\frac{\partial \Psi}{\partial T}\right)_q=\frac{(\partial q/\partial T)_{\Psi}}{(\partial q/\partial \Psi)_T}=\frac{\Psi(\partial C_m/\partial T)_{\Psi}}{C_m}
\label{eq:adiab2.3}
\end{equation}
with $q=C_m \Psi$. Thus, $(\partial \Psi/\partial T)_q$ is different from zero if the capacitance $C_m$ is temperature dependent. The capacitance is given by
\begin{equation}
C_m=\varepsilon_0\varepsilon \frac{A}{D}
\label{eq:adiab2.4}
\end{equation}
where $A$ is the area and $D$ the capacitor thickness. The capacitance will be temperature dependent if either the dielectric constant $\varepsilon$, the area $A$ or the thickness $D$ are temperature dependent. This is clearly the case for lipid membranes close to melting transitions since they display strong changes in area and thickness at the melting temperature \cite{Heimburg1998}. It has been shown that the capacitance can change by up to a factor of 2 in the melting transition \cite{Heimburg2012, Zecchi2017}. Adiabatic charging of a capacitor shall not be confused with the condenser theory described below in section \ref{thecondensertheory}.


\section{Application to nerve membranes}
\label{applicationtonervemembranes}

The following sections discuss the reversible heat production in nerves in the light of the thermodynamic considerations discussed in the previous section.


\subsection{The condenser theory}
\label{thecondensertheory}

The condenser theory is an attempt to rationalize the heat production in the Hodgkin-Huxley model. It has been discussed in various publications, e.g. in \cite{Abbott1973}. While the HH-model does not explicitly mention temperature and heat, one can try to translate the electrical elements (ion channels, batteries and capacitors) into a thermodynamic language.

In the condenser model, the ion concentrations of Na$^+$ and K$^+$ (the batteries) are treated in analogy with the expansion of ideal gases. The energy of an ideal gas only depends on temperature and is independent of pressure and volume. Since the change of energy is given by $dE=dQ+dW$, an isothermal expansion during which no work is performed does not involve any heat release or absorption. Fig.\,\ref{zeichnung} shows the discharging (depolarization) and recharging of the axonal membrane (depolarization) according to Hodgkin and Huxley. In step $b$, the energy of the capacitor is dissipated, which leads to a transfer of heat into the environment (it gets warmer). No work is performed ($dE=dQ$). In step $c$, work is performed by recharging the membranes while expanding the potassium gas across the membranes ($dW=-dQ$). In this step, heat is absorbed from the inside of the axion (the cytoplasm gets colder) and work is done upon recharging the capacitor. The charge on the capacitor in the end state $d$ is the same as in the beginning ($a$). Integrated over the total cycle, the amount of heat dissipated and reabsorbed adds up to zero. The magnitude of this heat exchange is solely dictated by the energy of the charged capacitor.

In the derivation of the Hodgkin-Huxley model, the capacitance was arbitrarily assumed to be constant. If we assume conservation of energy in eq. (\ref{eq:adiab2.1}) and further assume that the transmembrane voltage is independent of temperature, we obtain for the heat release in the first phase of the action potential
\begin{eqnarray}
dQ&=&-\Psi dq\quad\mbox{or} \quad \Delta Q= -\int_{q_0}^{q}\Psi dq \\
&=&-\frac{1}{2}C_m (\Psi_{max}^2-\Psi_0^2)=\frac{1}{2}C_m (V_{max}^2-V_r^2)\nonumber
\label{eq:condenser1}
\end{eqnarray}
where $\Psi_0$ corresponds to the resting potential $-V_{r}$. $\Psi_{max}$ is $-V_{max}$, which is the voltage at the peak of the nervous impulse. This calculation is called the `condenser theory'. If the absolute value of the resting potential is larger than that of the depolarized membrane, the heat exchange during the first phase of the action potential is negative (heat release) while it is positive during the second phase (heat absorption). This is qualitatively what has been found during the action potential.

Let us assume a biological membrane with a constant capacitance of $C_m=0.68$ \textmu F\slash cm$^2$. This corresponds to a DPPC membrane with a thickness of 3.92 nm, and a dielectric constant of $\varepsilon=3$ for its hydrophobic core. This value is close to those obtained for biological membranes (0.91 \textmu F\slash cm$^2$ for squid membranes \cite{Hodgkin1952b} and 0.5 \textmu F\slash cm$^2$ in mast cells \cite{Solsona1998}). The area of DPPC in its liquid state is 0.629 nm$^2$ per lipid, which yields a molar area for a bilayer of 189400 m$ ^2$. Charging this membrane from 0 to 100 mV according to eq. (\ref {eq:condenser1}) requires a work $\Delta W= 6.4$ J\slash mol. This is in fact a quite small energy. According to the condenser theory, this energy will be converted into the heat $\Delta Q=-\Delta W$ upon discharging the capacitor. This is the heat transfer measure according to the condenser theory.

Fig. \ref{heat_voltage_ritchie1985} shows that the heat release and re-uptake during the action potential are in fact proportional to the energy of a capacitor, i.e., it is proportional to the square of the voltage and is exactly in phase with voltage changes.

Whether the condenser theory can account for the observed heat response during the action potential has been the topic of several publications. In the first publication on this issue \cite{Abbott1958} it is stated that

\begin{quote}
	\emph{``The heats so calculated are of the right order of size, but on present evidence the time relations seem to be quite wrong.''}
\end{quote}

However, in \cite{Howarth1968}, the authors found that the condenser theory does not account for most of the measured heat. They state:

\begin{quote}
	\emph{``The condenser theory, according to which the positive heat represents the dissipation of electrical energy stored in the membrane capacity, while the negative heat results from the recharging of the capacity, appears unable to account for more than half of the observed temperature changes.''}
\end{quote}

The heat exchange recorded during the action potential is significantly larger than what would be possible if the nerve pulse simply consists of discharging and recharging of a capacitor. The authors concluded that

\begin{quote}
	\emph{``It seems probable that the greater part of the initial heat results from changes in the entropy of the nerve membrane when it is depolarized and repolarized.''}
\end{quote}

This is in agreement with a remark by Ritchie \cite{Ritchie1973} that

\begin{quote}
	\emph{``Part of the initial positive heat production is due to the free energy changes that accompany the discharging of the capacity; but most seems due to a large decrease in entropy in the dielectric on removal of the electric field across it. The observed changes in birefringence of the nerve on stimulation may agree with this interpretation.''}
\end{quote}

In view of these concerns we will explore alternatives to the the condenser theory.\\

\begin{figure*}[htbp]
	\centering
	\includegraphics[width=395pt,height=123pt]{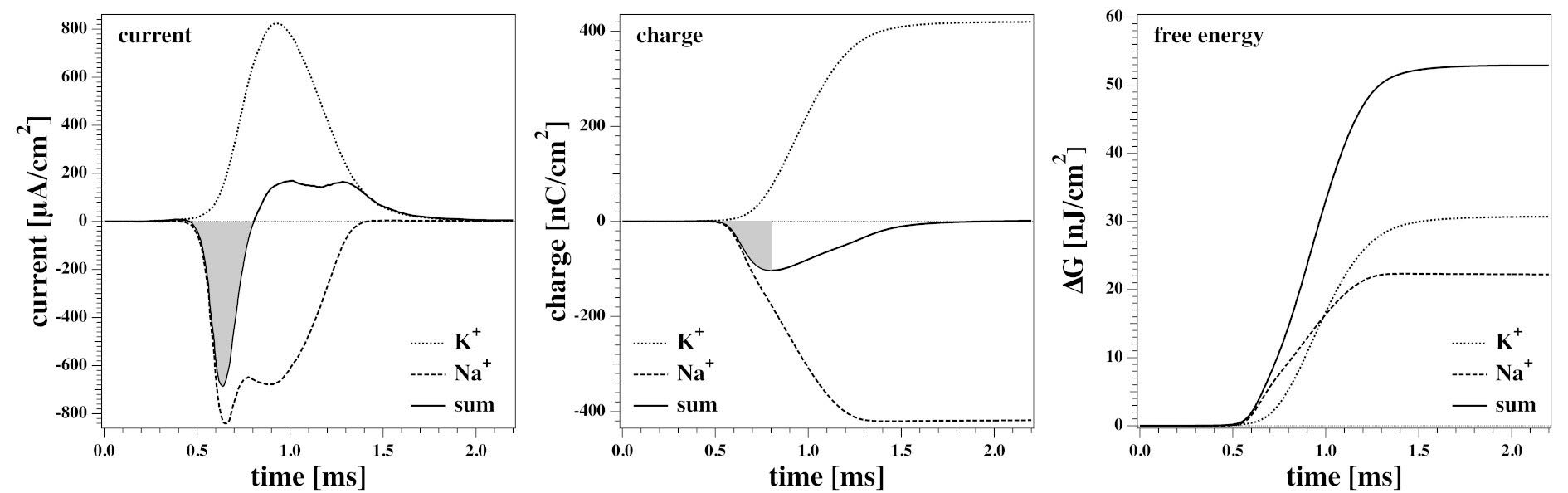}
	\caption{\small\textit {Left: Predicted potassium and sodium currents during the action potential in squid in the Hodgkin-Huxley model. The data were extracted from the original paper \cite{Hodgkin1952b}. Since potassium and sodium currents have opposite directions, their associated currents mostly compensate each other. Only the current in the grey-shaded region can be used to charge the membrane capacitor. Center: The integral over the currents yields the total number of charges flowing across the membrane. Only the sum of potassium and sodium charges is associated to charging the membrane. It adds to zero after the action potential. Right: Since the concentration differences and the Nernst potentials of potassium and sodium are known, one can calculate the free energy dissipation during the action potential. Since the free energy change display the same sign for potassium and sodium currents, the dissipation is substantial}}
	\label{figure_hh1952}
\end{figure*}

\subsubsection{Other ion-current based explanations}
\label{otherion-currentbasedexplanations}

The condenser theory is an attempt to understand the heat of nerves in the context of the Hodgkin-Huxley model that proposed currents through the membrane as the main origin of the action potential. All other phenomena are considered as epiphenomena that do not change the overall story.
In this context, there exist attempts to maintain the overall picture of Hodgkin and Huxley while deviating in technical detail. An interesting attempt was made by Margineanu et al. \cite{Margineanu1977} who considered the change of the free energy of ions in a field. If sodium and potation are considered as ideal gases charging the membrane, the capacitive energy is the only term contributing to changes in heat. However, if one considers the electrical energy of the ions in the field created by the action potential, i.e., the changes in electrochemical potential, according to \cite{Margineanu1977} one finds additional reversible contributions to the heat signature that also integrate to zero over the complete action potential because the membrane has the same potential after the action potential than before. The authors also added heat contributions due to free energy changes of ion channel proteins when they open and close. El Hady and Machta \cite{ElHady2015} maintained the original condenser model. The heat of charging the capacitor has been considered completely separate from the source of the voltage changes, i.e., the energy of the capacitor has been decoupled from the action potential whose origins are not discussed in the paper. We believe that this is not an appropriate manner to analyze the thermodynamics of a nerve in a self-contained manner because one cannot consider the origins of the voltage change independent from the energy of the capacitor. Lichtervelde et al. \cite{Lichtervelde2020} provided a recent attempt to look at the condenser model in a very interesting and more sophisticated thermodynamic manner. Besides considering the capacitor only, they also considered contributions of fixed surface charges, i.e., a membrane polarization. It is known that biological membranes possess permanent polarizations (discussed in \cite{Mosgaard2015b}). It should be noted though that in our opinion additional assumptions also constitute a deviation from the Hodgkin-Huxley model in which the capacitance is fixed and $d C_m/dt$-terms are omitted.

\subsubsection{Dissipation of free energy in the Hodgkin-Huxley model}
\label{dissipationoffreeenergyinthehodgkin-huxleymodel}

The ionic fluxes proposed in the Hodgkin-Huxley (HH) model represent a very inefficient process for creating an action potential. Fig. \,\ref{figure_hh1952} (left) shows the time course of sodium and potassium currents calculated by Hodgkin and Huxley in their seminal paper \cite{Hodgkin1952b} (convention: current from inside to outside is positive). One finds that they flow in opposite directions with significant temporal overlap. Only the sum of these currents is the ionic current contributing to charging the capacitor (grey-shaded region). The center panel of this figure shows the integral of these currents, i.e., the total charge transported from inside to outside. The free energy change of each ion following it's gradient from one side of the membrane to the other is given by $\Delta \mu_K=kT\ln([\mbox{K}_{in}]/[\mbox{K}_{out}])$ for potassium and $\Delta \mu_{Na}=kT\ln([\mbox{Na}_{out}]/[\mbox{Na}_{in}])$, which has to be multiplied by the number of ions that flow across the membrane given in the center panel of Fig. \,\ref{figure_hh1952}. We take the textbook value $[\mbox{K}_{in}]=400$ mM, $[\mbox{K}_{out}]=20$ mM, $[\mbox{Na}_{in}]=50$ mM and $[\mbox{Na}_{out}]=440$ mM \cite{Johnston1995}. The total dissipated free energy per cm$^2$ is shown in Fig. \,\ref{figure_hh1952} (right). The maximum of the solid line yields $\Delta F=$ 52.9 nJ\slash cm$^2$ at T=10$^\circ$C which indicates the total free energy dissipated by ions during the action potential.

The membrane changes its potential by 110 mV relative to the resting potential. The capacitance is assumed being $C_m=$1 \textmu F$/$cm$^2$. This yields an energy of the charged capacitor at the maximum of the action potential of $E= \frac{1}{2}C_m V_{max}^2 = 6.05$ nJ \slash cm$^2$ which is the nerve signal transported from one end to the other end of the axon.  This is just 11.4\% of the dissipated free energy over the length of one action potential, which has to be multiplied by the overall length of the axon. The Hodgkin-Huxley model therefore dissipates about 10 times more free energy than necessary to charge the capacitor over the length of one action potential. This happens over each length of the action potential. Additionally, the energy of the capacitor is not stored but has to be recreated by ion flows at any given position of the axon. The velocity of the action potential in the squid axon was calculated to be 18.8 m\slash s and lasts about $\Delta t =1.3$ ms, which yield a length of the pulse of $\Delta x=v\cdot \Delta t$ = 2.5 cm. For a nerve that is 25 cm long (10 lengths of the action potential), 100 times more free energy is dissipated compared to the energy of the signal. Thus, the Hodgkin-Huxley action potential is a very inefficient mechanism that dissipates 2 orders of magnitude more free energy than what corresponds to the signal that is transported. Since the ions have to be pumped back by NaK-ATPases, one expects a very large metabolic heat dissipated on the timescale of the action of the pump due to ATP-hydrolysis. If 50\% of the free energy stored in the ATP were dissipated as heat, one expects the dissipation of about 5 times more metabolic heat over the length of a single action potential than the heat that is due to charging the capacitor in the condenser model. This is clearly not the case in Fig. \,\ref{abbott1958}C from the seminal paper of \cite{Abbott1958}, where the drift in the baseline (which may represent metabolic heat) is roughly 38\% of the initial heat, i.e., it is more than one order of magnitude smaller. Later publications found a much lower or even absent drift in the baseline (e.g., \cite{Ritchie1985}), which may indicate the complete absence of any metabolic heat on the timescale of an action potential. However, this is not expected for the action of ATPases that has to accompany the nervous impulse in the HH-model.


\subsection{Heat release in the soliton model}
\label{heatreleaseinthesolitonmodel}

Biological membranes exist in a state slightly above a melting transition \cite{Muzic2019}. In a lipid melting transition, both enthalpy and entropy change \cite{Heimburg2007a}. In the soliton model, the nerve pulse consists of region in an ordered lipid state embedded in the liquid environment of the resting nerve membrane. Thus, during the nerve pulse the entropy in the membrane changes, which is accompanied by a transient release of latent heat into the nerve environment.

We will determine the temperature increment due to adiabatic compression close to the melting transition in biomembranes. According to eqs. (\ref{eq:adiabmem3}) and (\ref{eq:adiabmem4}), the temperature increment is given by
\begin{equation}
dT = -\frac{T}{c_A}\left(\frac{\alpha_\Pi}{\kappa_T^A}\right) dA
\label{eq:adiabsoliton1}
\end{equation}
We have previously shown that in lipid transitions the thermal area expansion coefficient is proportional to the heat capacity \cite{Heimburg1998}:
\begin{equation}
\alpha_\Pi=\left(\frac{\partial A}{\partial T}\right)_\Pi=\gamma_A \left(\frac{\partial H}{\partial T}\right)_\Pi=\gamma_A c_\Pi \;.
\label{eq:adiabsoliton2}
\end{equation}
Here, we used a coefficient $\gamma_A=0.89$ m\slash N that is of similar order in the lipid DPPC, biological lung surfactant extracts, and in bacterial and nerve membranes \cite{Ebel2001, Muzic2019}: Here, $c_{\Pi}$ is the heat capacity of the lipids at constant lateral pressure. Further, we have shown that the isothermal area compressibility, $\kappa_T^A$, is also proportional to the heat capacity:
\begin{equation}
\kappa_T^A=\gamma_A^2 T c_\Pi
\label{eq:adiabsoliton3}
\end{equation}
Inserting this into eq. (\ref{eq:adiabsoliton1}) yields
\begin{equation}
dT =-\frac{1}{c_A}dH
\label{eq:adiabsoliton4}
\end{equation}
Here, $c_A$ is the heat capacity of the heat reservoir that includes the interfacial water. If compression is infinitely slow, the whole water volume may be involved and the heat capacity is very large. Consequently, the temperature increment will be very small. The heat transferred upon the compression of the membrane is given by
\begin{equation}
c_A \Delta T = -\int_{A_f}^{A}dH =+\Delta H \;,
\label{eq:adiabsoliton5}
\end{equation}
where $\Delta H$ is the latent heat of the membrane transition. The reference state is the membrane in its fluid state with area $A_f$, and the area A is the membrane area at the maximum of the electromechanical pulse in the nerve membrane. The maximum pulse amplitude corresponds to the minimum membrane area, which is the area of the gel state membrane, $A_g$.

The latent heat of the artificial lipid DPPC is 35 kcal\slash mol. In section \ref{thecondensertheory} we have calculated that the heat release upon discharging the membrane capacitor in the condenser theory is about 6.4 J\slash mol. Thus, the possible heat transfer in the soliton model (or any other model involving phase transitions in membranes) could be up to 5000 times larger than in the purely electrical view.


\subsection{Underestimation of the heat transfer from axons in experiments}
\label{underestimationoftheheattransferfromaxonsinexperiments}

\begin{figure}[htbp]
	\centering
	\includegraphics[width=225pt,height=114pt]{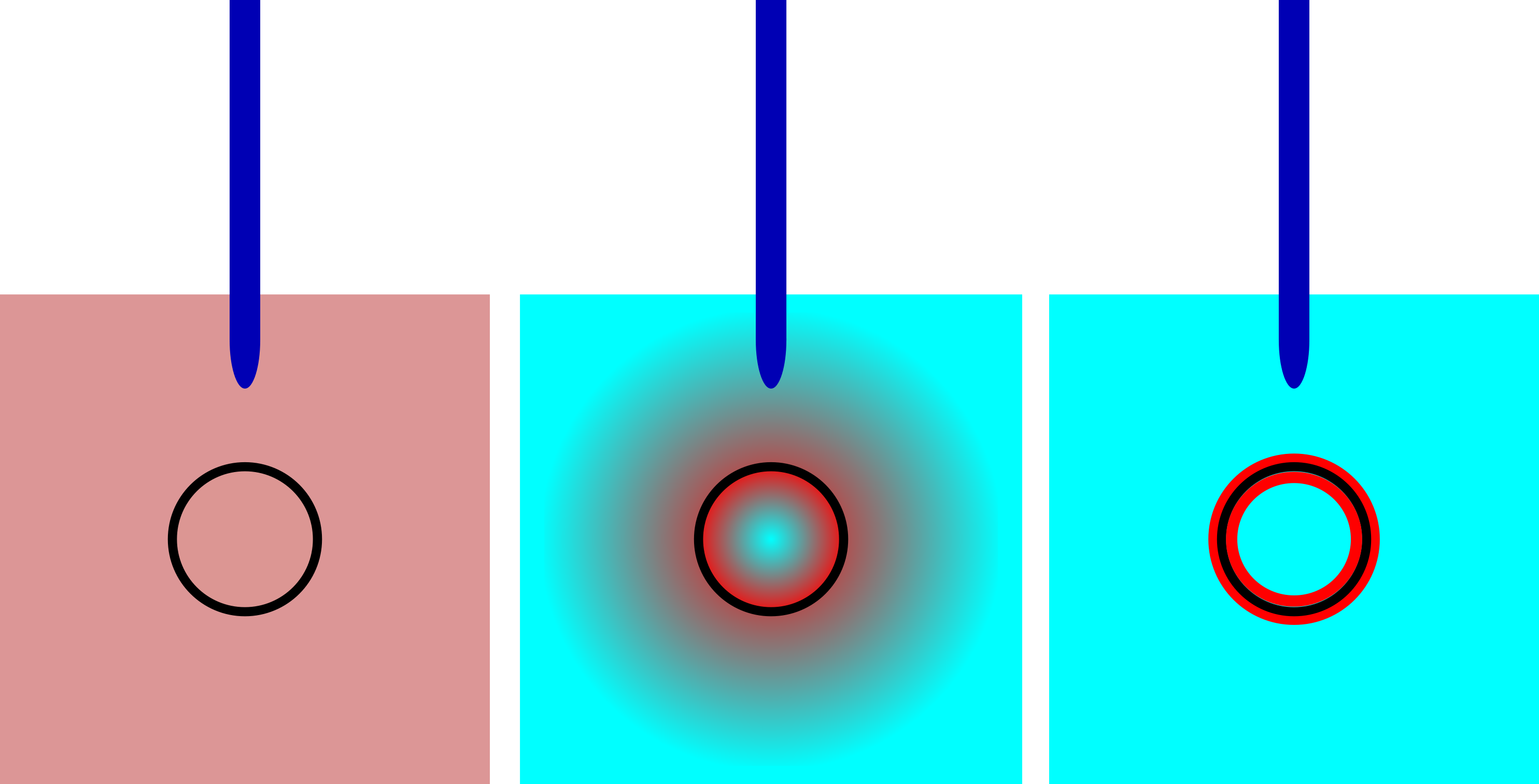}
	\caption{\small\textit {Three different situations for heat release around a nerve cylinder (black circle) as measured by a thermocouple (blue bar). The red-blue color scale schematically describes the temperature distribution during the action potential. Left: All heat released from the nerve membrane diffuses fast in the medium and the temperature is uniform. Center: Heat diffuses slowly into the medium and only parts of the heat transfer are seen by the thermocouple. Right: The heat remains in the interfacial aqueous layer surrounding the axon. A measurement with a thermocouple largely underestimates the total heat release.}}
	\label{heat_nerve2}
\end{figure}

In the experiments described above, temperature changes were measured with thermocouples or temperature-sensitive resistors. Howarth and collaborators (e.g., \cite{Howarth1968}) compared the temperature change recorded during an action potential with the temperature change caused by an artificial externally applied heat pulse of known magnitude (heat block analysis). In order to calculate the total heat released during the action potential, the authors assumed that the temperature is uniform in the nerve tissue and the cytosol, i.e., that heat diffuses and equilibrates very fast. The heat capacity is assumed to be that of free water, i.e., about 1 cal\slash cm$^3$K. This is not necessarily a realistic assumption because during the action potential the heat might not diffuse from the nerve membrane into the aqueous volume, for instance because the direct interface of the membrane might consist of structured water or molecules with a very different heat capacity. A thermocouple located in the aqueous medium might see a temperature change much smaller than what is expected at the surface of the membranes. This leads to an underestimation of the measured heat production. These cases were considered theoretically in \cite{Mosgaard2013a}.
The left hand panel of Fig. \ref{heat_nerve2} depicts a situation where heat conduction is fast and the temperature is uniform. In such an situation, the translation between temperature change and the heat release is simple because the heat capacity is just that of the total aqueous volume, and $\Delta Q=c_A \Delta T$. Fig. \ref{heat_nerve2} (center panel) shows a situation in which the temperature is not uniform but larger close to the neuron due to incomplete equilibration. In such a situation, the heat transfer is underestimated if the situation in the left hand panel is taken as the basis for a calculation. The right hand panel of Fig. \ref{heat_nerve2} depicts a situation in which the heat is transferred into the interfacial layer only, and the thermocouple sees no temperature change at all. One might falsely assume that no heat transfer takes place. Therefore, the experimentally determined transfer of heat from the axonal membrane into the medium is just a lower estimate based on the assumption of fast heat conduction.

In ultrasound experiments with lipid membrane vesicles at a frequency of 5 MHz (corresponding to a timescale of 0.2 \textmu s), the sound velocity of the lipid dispersion is well predicted when assuming that the membrane does not transfer any heat into the water upon compression \cite{Halstenberg1998, Schrader2002}. In order to calculate the adiabatic compressibility, it is sufficient to assume that all heat is absorbed by the lipid chains and no exchange with the aqueous medium takes place. This indicates that heat transfer from membranes into the environment may be much slower than expected from the thermal conductivity of plain water. It must occur on a time scale much longer than 0.2 \textmu s as judged from 5 MHz ultrasound experiments. Therefore, it is likely that the heat transfer between neuronal membranes and its environment is underestimated in the interpretation of experiments and that the thermocouples may not be able to detect most of the heat release.
Slow action potentials imply a more complete heat exchange. This results in a larger compressibility which in a circular argument leads to a slower action potential. Therefore, speed and magnitude of heat exchange are coupled. This was discussed previously in \cite{Mosgaard2013a}. This could represent another reason why myelinated nerves which don't allow for a complete heat exchange display a larger velocity.

In a biological membrane, the heat sink contributing to the heat capacity $c_A$ may include everything that can absorb the heat originating from changes in lipid conformation. This may comprise vibrations of the hydrocarbon chains, bulk water, but also membrane proteins and interfacial ordered water. If the interface has a high heat capacity, a temperature measurement in the bulk water might largely underestimate the heat release from the action potential. Interestingly, Tasaki always favored a view in which the superficial protein layer undergoes a phase transition during the action potential \cite{Tasaki1999}. A phase transition in the proteins would imply a large heat capacity in the superficial protoplasmic layer.

\section{Discussion}
\label{discussion}

In this article we reviewed and analyzed the striking heat exchange during the action potential, which has occupied the minds of excellent scientists for more than 170 years. The present state-of-the-art of the experimentation indicates that during the action potential in nerves an initial phase of heat release is followed by one of heat reabsorption. Within experimental accuracy, no heat is dissipated during the action potential \cite{Abbott1958}. This indicates that no metabolism occurs during the nerve pulse. The integrated rate of heat exchange is in phase with the action potential and proportional to the voltage squared. The overall heat changes are larger than what can be explained by a purely electrical theory. The heat changes were found in crustaceans, in fish, and in mammalian nerves. Thus, this effect seems to represent a generic property of nerves.

The measurement of temperature and heat changes in nerves is exceptionally difficult. The observed changes range from a few \textmu K to a few 100 \textmu K. A standard K-type chromel-alumel thermocouple generates a voltage of 41 \textmu V\slash K, which corresponds to 0.41 nanovolt during an action potential with $\Delta T=10$\textmu K. This is on the very edge of what can be detected. Therefore, in most studies thermocouples were combined, e.g., 96 in a thermopile \cite{Abbott1958}. This increases the signal to about 40 nV. Combining thermocouples increases their mass and their heat capacity. Therefore, the temperature sensor has a time constant, which in the above case was larger than 60ms. In order to measure temperature changes, measurements have focused on action potentials of nerve bundles at low temperatures, which display very long action potentials of the order of 100--200 ms.
However, action potentials under physiological conditions are fast - in most cases of the order of 1--2 milliseconds. This cannot be detected by thermopiles. For this reason, Tasaki and Iwasa \cite{Tasaki1981} used a temperature-sensitive polymer film, which reduced the time constant to a few milliseconds.

The smallness of the temperature effect may lead to the tempting prejudice that the heat changes are small and unimportant. However, when it comes to understanding magnitudes intuition can be misleading. Whether we consider an effect small or big is often guided by the sensitivity of our instrumentation. Voltage is easy to measure and therefore we assume that the nerve pulse is dominated by voltage changes. We have shown above that a small temperature change can be associated with energy changes that are bigger than the electrical energy of a charge axion. From the perspective of an energy, the heat change is in fact larger than the electrical effect, even if temperature changes are small. \\

The reversible heat production has puzzled all authors of heat studies and many different attempts have been made to explain this phenomenon. The following origins have been discussed:

\begin{enumerate}
	\item Joule heating by action currents \cite{Abbott1973}. Joule heating originates from electrical currents through electrical resistors. In the case of the nerve membrane, it would be accompanied by a cooling of the batteries that drive the currents (the concentration differences of ions across the membrane).
	
	\item Thermoelectric temperature changes \cite{Abbott1973}. This refers to the Peltier effect that states that voltage gradients can lead to temperature differences and heat flux.
	
	\item Heat of ionic interchange \cite{Abbott1973, Howarth1975, Tasaki1989}. When two solutions containing different ions are mixed, one can measure a heat of mixing which can be both positive and negative. The heat of ion mixing would be zero for ideal gases and ideal solutions, which is assumed for the derivation of membrane-potentials in the Hodgkin-Huxley model.
	
	\item Condenser theory \cite{Abbott1958, Howarth1968, Abbott1973, Tasaki1989}. This theory considers the energy and heat changes associated to charging and discharging the membrane capacitor.
	
	\item Reversal of sodium pump \cite{Abbott1973}. This theory considers mechanisms for ion translocation across the nerve membrane that go beyond the Hodgkin-Huxley model and require extra molecular elements in a membrane theory.
	
	\item Cooperative interaction between the ion-exchange properties of the superficial layer of the axoplasm and the nerve membrane \cite{Tasaki1989}. Here, it is assumed that the changes during the action potential cause the dissociation or association of ion on the interface of the nerve membrane such as in an ion exchanger.
	
	\item Conformational changes of channel proteins during the action potential accounting for entropy changes\cite{Margineanu1977}
	
	\item Reversible melting of the nerve membrane other change in membrane entropy \cite{Howarth1968, Abbott1973, Howarth1975, Heimburg2005c}. Here, it is considered that there exist changes in the physical structure of the membrane as a whole such as melting of membrane lipids, or the change in the dielectric properties on the membrane that would be related to ordering and disordering of dipoles within the membrane.
	
\end{enumerate}

Most of these possibilities above are of purely speculative nature which were dismissed because one could not provide a natural explanation of the magnitude of the heat change and why they should be in phase with voltage changes. The two possibilities that were taken most seriously and discussed quantitatively are the condenser theory and the entropy changes in the membrane. Howarth et al. \cite{Howarth1975} state that

\emph{``The positive heat seems to be derived from two sources. First, there is a dissipation of the free energy stored in the membrane capacity. Secondly, there is an evolution of heat corresponding with a decrease in entropy of the membrane dielectric with depolarization.''}

The condenser theory assumes that the heat originates from discharging and recharging of the axonal membrane during the action potential (discussed in section \ref{thecondensertheory}), which is the main element of the Hodgkin-Huxley model \cite{Hodgkin1952b}. It is described in section \ref{thehodgkin-huxleymodel}. However, it is difficult to reconcile the magnitude of the heat changes with a purely electrical theory. The heat exchange is too small \cite{Howarth1968, Abbott1973, Howarth1975}, and it is not agreed whether the shape of the heat release is theoretically superimposable with voltage changes \cite{Abbott1958, Tasaki1989} - as is the case with the experimental data. For this reason, the condenser theory has been dismissed by most authors because the measured heat changes are significantly larger than what could be explained by charging of a capacitor using known capacitance values and using the voltage changes that have been measured. Nevertheless, any thermodynamic theory that considers voltage changes during the nerve pulse and acknowledges that membranes have a capacitance must include the effects of the condenser (capacitor). However, an adiabatic pulse would not be described by the condenser theory but rather by the adiabatic charging of a capacitor as described in section \ref{electricalsystems:adiabaticchargingofacapacitor}.

The concept that membrane entropy might alter during the action potential was first stated by \cite{Howarth1968}

\begin{quote}
	\emph{``It seems probable that the greater part of the initial heat results from changes in the entropy of the nerve membrane when it is depolarized and repolarized.''}
\end{quote}

and \cite{Abbott1973}:

\begin{quote}
	\emph{``Another idea that has received attention is that the open-circuit cooling has its origin in the membrane itself. If the membrane change involved a structural change in the lipid fraction analogous to the melting of a fat it would be accompanied by an absorption of heat (viz. the latent heat of fusion). {\ldots} Certainly the membrane undergoes structural changes during the action potential that may well be accompanied by detectable heat changes {\ldots} and may indeed by analogous to the melting of some membrane fraction {\ldots}.''}
\end{quote}

It has to be noted that at the time of this statement, the physical chemistry of lipid membranes was in its infancy, and melting transitions in biological membranes had just been observed. The existence of such transitions slightly below physiological temperature in biological membranes and nerves is well documented today (e.g., \cite{Muzic2019}).

Experiments show that the nerve membrane contracts during the nervous impulse \cite[]{Wilke1912b, Tasaki1989}, which indicates that the surface area of the nerves alter during the action potential. If one assumes that there exist adiabatic area changes during the action potential, one can derive the temperature changes in the medium surrounding the nerve membranes by
\begin{equation}
dQ=c_A dT = -T\left(\frac{\alpha_\Pi}{\kappa_T^A}\right) dA\;,
\label{eq:discuss1}
\end{equation}
i.e., it depends on the heat capacity $c_A$ of the medium surrounding the nerve, the thermal area expansion expansion coefficient $\alpha_{Pi}$ of the membrane and the isothermal area compressibility $\kappa_T^A$ (see section \ref{membranes}).

When the membrane is pushed through its transition by reducing the area, the heat release corresponds to the latent heat (enthalpy) of the transition (section \ref{heatreleaseinthesolitonmodel}). It is reabsorbed when going back to the original state because the enthalpy is a function of state. If one represents the action potential as an adiabatic pulse, i.e., an electromechanical soliton, no net heat is produced \cite{Heimburg2005c}. In the soliton model (section \ref{heatreleaseinthesolitonmodel}), the membrane is reversibly shifted through a melting transition in the membrane. Heat is reversibly moved from the nerve membrane into the surrounding medium and back. The magnitude corresponds to the latent heat of the transition. The heat translocations are in phase and proportional to voltage changes. The effect is analogous the adiabatic heating and cooling of gases and springs during oscillations - but amplified due to the presence of a transition. We have shown that the heat release due to the mechanical changes is orders of magnitude larger that the electrical energy of a membrane capacitor. Therefore, it is quite plausible to assume that reversible melting of the membrane can partially or fully explain the measured heat changes. Heat transfer is predominantly into the external environment of the membrane and back. Heat transfer along the membrane would just give rise to the propagation of the wave meaning that upon dislocation of the soliton heat is transferred along the membrane without any heating of the environment. This is a general feature of adiabatic waves.

The Hodgkin-Huxley model \cite{Hodgkin1952b} is an electrical theory of the nerve pulse. As discussed above, all authors involved in measuring heat changes concluded that the heat exchange during the action potential is too large for the reversible charging of a capacitor. Several of them proposed to consider reversible changes in membrane entropy. This is exactly the central point in the soliton model, in which the action potential consists of a electromechanical pulse reminiscent of sound. In the soliton model, the heat originates from the latent heat of a melting transition in the nerve membrane. As discussed in section \ref{heatreleaseinthesolitonmodel}, the heat release from a soliton may be up to 5000 times larger than the energy of the membrane capacitor. Thus, while the the measurements prove that a purely electrical theory cannot account for the measured heat, it is presently unclear why the heat is not much larger as required in the thermodynamic view. In section \ref{underestimationoftheheattransferfromaxonsinexperiments} we argued that the assumption of complete temperature equilibration across the aqueous medium of nerve tissue may just be false (cf., \cite{Mosgaard2013a}). We have shown that the relaxation times of transitions in nerve membranes can be more than 100 ms for typical biological membranes \cite{Grabitz2002, Seeger2007} and based on ultrasonic experiments one has too conclude that heat transfer from membranes into the aqueous environment might be much slower than expected \cite{Halstenberg1998}. This would imply that the measured change in temperature may largely underestimate the heat transfer. But it is clear that in the solitonic view one would expect a much larger heat release and that the magnitude of the heat exchange in both condenser theory and soliton model is far from being completely understood.

An isentropic process is a thermodynamic process that is both adiabatic and reversible. The condenser model is adiabatic but not reversible, while the soliton model is both adiabatic and reversible – hence it is isentropic. The first process is dissipative, the second is not. In the past, various authors have proposed sound based approaches. Already Helmholtz noted that the nerve pulse displays a velocity close to the speed of 2D-sound in soft matter \cite{Helmholtz1852}. Wilke also discussed the propagation of mechanical waves \cite{Wilke1912a, Wilke1912b}. Hill and Bayliss argued that the signature of nerve pulses might involve reversible physical or physicochemical changes comparable to physical waves \cite{Hill1912, Bayliss1915}. Thus, the idea to treat the nerve pulse as a reversible thermodynamic event has been around for more than 100 years. Kaufmann reiterated in 1989 that the nerve pulse might be of reversible thermodynamic nature \cite{Kaufmann1989e} in which entropy is conserved. He suggested that the pulse is related to transitions in the nerve membrane. This is also the basic idea in the soliton model, which is the first detailed realization of these ideas. \cite{Heimburg2005c, Andersen2009, Lautrup2011}. The solitonic nature of the pulse explains how the nerve pulse maintains its shape over long distances. We have shown here that the representation of the action potential as an adiabatic pulse is consistent with the measured data. In particular, the reversible heat exchange in a model involving phase transitions can be of a magnitude that is considerably higher than the energy of charging a capacitor. This is consistent with the fact that one finds mechanical changes in thickness and length of the nerve of suitable magnitude (section \ref{otherimportantthermodynamicfindingsonnerves}). An adiabatic action potential suggests that there is little friction and no loss in free energy during the action potential. In contrast, the loss in free energy during the Hodgkin-Huxley pulse can be significant and exceed the energy of the charged capacitor (the signal) by several hundred times (section \ref{dissipationoffreeenergyinthehodgkin-huxleymodel}. The electrical theory is in fact an extremely wasteful uneconomic way to send signals along a nerve fiber.

In a booklet from 1990, Schoffeniels and Margineanu wrote \cite{Schoffeniels1990}

\begin{quote}
	\emph{''{\ldots} one is faced with the not so common situation that some undisputed experimental results, namely those concerning the heat evolved by the stimulated nerves, stubbornly resisted the attempts to integrate them in the otherwise successful ionic theory of excitation. The typical human behaviour was to almost forget those results. ''}
	They quote Hill \cite{Hill1965}: 
	\emph{``nerve heat production is rather a nuisance; things would be so much simpler without it.''}\\
	and comment: 
	\emph{``What he meant was that heat production prevents the nerve to function in ideal conditions, but his remark would be equally suited with respect to the theories about nerve impulse. {\ldots}. No explanation of a natural phenomenon can be accepted until it copes with the laws of thermodynamics.''}
\end{quote}

For the above reasons, there can be no doubt that the thermal changes found in nerves are a very important observation. It would be highly desirable if the heat experiments on nerve were reproduced and refined in order to describe the heat exchange during the fast action potentials in nerves and single neurons with short pulse duration. In particular, it would be of considerable importance if one could extend the experiments to fast action potentials on a time scale of 1ms, which so far has not been possible due to the long response time of the present instrumentation. Such experiments are absolutely necessary if one wants to understand nerves on a fundamental physical basis. It is very likely that more detailed experiments will provide important insights into a thermodynamic understanding of the nerve pulse.

\vspace{0.5cm}
\textbf{Acknowledgment:} I am grateful to Prof. Andrew D. Jackson for critical reading of the manuscript, helpful suggestions and for intense discussions. The final quote by Schoffeniels and Martineanu was brought to my attention by Shamit Shrivastava (Oxford).

\small{

}

\end{document}